\documentclass[11pt,letterpaper,onecolumn]{article}
\usepackage{graphicx}
\usepackage{subfigure}
\usepackage{url}
\usepackage{color}
\usepackage{multirow}
\usepackage{amssymb,amsmath}
\usepackage{algorithm,algorithmic}
\usepackage{cite}
\usepackage{setspace}

\newcommand{\SF}{{\cal F}}
\newcommand{\SX}{{\cal X}}
\newcommand{\SY}{{\cal Y}}

\newtheorem{theorem}{Theorem}
\def\done{\hspace*{\fill}\rule{1.8mm}{2.5mm}\\}

\begin{document}

\title{CORE: Augmenting Regenerating-Coding-Based Recovery for Single
and Concurrent Failures in Distributed Storage Systems}

\author{Runhui Li, Jian Lin, Patrick Pak-Ching Lee\\
Department of Computer Science and Engineering, The Chinese University of Hong
Kong\\
{\it \{rhli, jlin, pclee\}@cse.cuhk.edu.hk}}

\maketitle
\thispagestyle{empty}

\begin{abstract}
Data availability is critical in distributed storage systems, especially when 
node failures are prevalent in real life.  A key requirement is
to minimize the amount of data transferred among nodes when recovering the 
lost or unavailable data of failed nodes.  This paper explores recovery
solutions based on regenerating codes, which are shown to provide
fault-tolerant storage and minimum recovery bandwidth.  Existing optimal
regenerating codes are designed for single node failures.  We build a system
called CORE, which augments existing optimal regenerating codes to support a
general number of failures including single and concurrent failures.  We
theoretically show that CORE achieves the minimum possible recovery bandwidth
for most cases.  We implement CORE and evaluate our prototype atop a Hadoop
HDFS cluster testbed with up to 20 storage nodes.  We demonstrate that our
CORE prototype conforms to our theoretical findings and achieves recovery
bandwidth saving when compared to the conventional recovery approach based on
erasure codes. 
\end{abstract}

%\begin{keywords}
%regenerating codes, failure recovery, distributed storage systems, coding
%theory, experiments and implementation
%\end{keywords}

{\bf Notes:} A 6-page shorter conference version of this paper
appeared in Proceedings of the 29th IEEE Conference on Massive Data Storage
(MSST), May 2013 \cite{li13}. 

\section{Introduction}

To provide high storage capacity, large-scale distributed storage systems have
been widely deployed in enterprises, such as Google File System
\cite{ghemawat03}, Amazon Dynamo \cite{decandia07}, and Microsoft Azure
\cite{calder11}.  In such systems, data is striped across multiple nodes (or
servers) that offer local storage space.  Nodes are interconnected over a
networked environment, in the form of either clustered or wide-area settings.  

%Data may be permanently lost due to disk crashes \cite{schroeder07}.
%However, it is also common to see short-term transient failures where storage
%nodes are temporarily unavailable.  Transient failures are prevalent not only
%in wide-area storage systems due to node outages or network partitions
%\cite{bhagwan04,chun06,nath06}, but also in clustered storage systems due to
%reboots or upgrades \cite{ford10,huang12}.  Thus, maintaining data
%availability against failures becomes critical in distributed storage
%systems.  

Ensuring data availability in distributed storage systems is critical, given
that node failures are prevalent \cite{ghemawat03}.  Data
availability can be achieved via erasure codes (e.g., Reed-Solomon codes
\cite{reed60}), which encode original data and stripe encoded data across
multiple nodes.  Erasure codes are defined by parameters $(n,k)$ (where
$k<n$), such that if any subset of $n-k$ out of $n$ nodes fails, the original
data remains accessible by decoding the encoded data stored in other $k$
surviving nodes.  Erasure codes can tolerate multiple failures, while
incurring less storage overhead than replication.

In addition to tolerating failures, another crucial availability requirement
is to {\em recover} any lost or unavailable data of failed nodes.  Recovery
is performed in two scenarios: (i) when the failed nodes are crashed and the
permanently lost data need to be restored on new nodes, and (ii) when the
unavailable data needs to be accessed by clients before the failures are
restored.  The conventional recovery approach, which applies to {\em any}
erasure codes, first reconstructs all original data to obtain the
lost/unavailable data.
Since the lost/unavailable data usually accounts for only a fraction of
original data, previous studies explore how to optimize the recovery
performance by minimizing the amount of data involved.  One class of
approaches is to minimize I/Os (i.e., the amount of data read from disks) 
based on erasure codes (e.g.,
\cite{huang12,khan12,sathiamoorthy13,wang10,xiang11}).
Another class of approaches is to minimize the recovery bandwidth 
(i.e., the amount of data transfer over a network during recovery) based on
{\em regenerating codes} \cite{dimakis10}, in which
each surviving node encodes its stored data and sends encoded data for
recovery.  In the scenario where network capacity is limited,
minimizing the recovery bandwidth can improve the overall recovery
performance.   In this work, we focus on exploring the feasibility of
deploying regenerating codes in practical distributed storage systems.

However, most existing recovery approaches, including those for minimizing
I/Os and bandwidth, are restricted to {\em single failure} recovery.  Although
single failures are common, node failures are often correlated and
co-occurring in practice, as reported in both clustered storage (e.g.,
\cite{ford10,schroeder07}) and wide-area storage (e.g.,
\cite{chun06,haeberlen05,nath06}).  To provide tolerance against 
{\em concurrent} (multiple) failures,  data is usually protected with a high
degree of redundancy. For example, Cleversafe \cite{cleversafe}, a commercial
wide-area storage system, use (16,10) erasure codes (i.e., up to 6 out of 16
concurrent failures are tolerable) \cite{plank09}.  Some wide-area storage
systems such as OceanStore \cite{kubiatowicz00} and CFS \cite{dabek01} employ
erasure codes with even higher double redundancy $(n,n/2)$.  We
believe that in addition to providing fault tolerance, minimizing the recovery
bandwidth for concurrent
failures will provide additional benefits for today's large-scale distributed
storage systems.   In addition, concurrent failure recovery is beneficial to
delaying immediate recovery \cite{bhagwan04}.  That is, we can perform
recovery only when the number of failures exceeds a tolerable limit.  This
avoids unnecessary recovery should a failure be transient and the data be
available shortly (e.g., after rebooting a failed node).  Given the importance
of concurrent failure recovery, we thus pose the following questions: (1) Can we
achieve bandwidth saving, based on regenerating codes, in recovering a general
number of failures including single and concurrent failures?  (2) If we can
enable regenerating codes to recover concurrent failures, can we seamlessly
integrate the solution into a practical distributed storage system?  

In this paper, we propose a system called CORE, which
supports both single and \underline{co}ncurrent failure \underline{re}covery
and aims to minimize the bandwidth of recovering a {\em general} number of
failures.  CORE augments existing optimal regenerating codes (e.g.,
\cite{rashmi11b,suh11}), which are designed for single failure recovery, to
also support concurrent failure recovery.  A key feature of CORE is that it
retains existing optimal regenerating code constructions and the underlying
regenerating-coded data.  That is, instead of proposing new code
constructions, CORE adds a new recovery scheme atop existing regenerating
codes. Our idea is to treat all but one failed nodes as logical surviving
nodes.   CORE first reconstructs the ``virtual'' data to be generated by those
logical surviving nodes.  By combining the virtual data with the real data
being generated by the real surviving nodes, CORE then reconstructs the
remaining failed node using existing optimal regenerating codes.  We apply the
same idea for all failed nodes. 

%achieves bandwidth saving over the conventional approach in reconstructing
%concurrent failures.  A side benefit is that it also allows the storage
%system to defer immediate recovery of single failures but instead
%reconstruct concurrent failures in batch.  

In summary, the contributions of this paper are three-fold. 
\begin{itemize}
\item
{\bf Theoretical analysis.} We theoretically show that CORE achieves the
minimum bandwidth for a majority of concurrent failure patterns.  We also 
propose extensions to CORE to achieve sub-optimal bandwidth saving even for
the remaining concurrent failure patterns.  Our analytical study validates
that CORE can recover concurrent failure patterns with significant bandwidth
saving over conventional recovery based on erasure codes. For example, for
(20,10), the bandwidth savings are 36-64\% and 25-49\% in the optimal and
sub-optimal cases, respectively.  We also show via reliability analysis that
CORE has significantly longer mean-time-to-failure (MTTF)
than conventional recovery. 
\item
{\bf Implementation.} We implement a prototype of CORE and demonstrate the
feasibility of deploying CORE in a practical distributed storage system.  As a
proof of concept, we choose the Hadoop Distributed File System (HDFS)
\cite{shvachko10} as a starting point.  CORE sits as a layer atop HDFS
and supports recovery for a general number of failures.  We build CORE
atop HDFS by modifying the source code of HDFS and its erasure coding
extension HDFS-RAID \cite{hdfsraid}.  We also adopt a pipelined
implementation that parallelizes and speeds up the recovery process.  
\item
{\bf Experiments.} We experiment CORE on an HDFS testbed with up to 20 storage
nodes.  Our experiments take into account a combination of different
factors including network bandwidth, disk I/Os, encoding/decoding overhead.
We justify that minimizing bandwidth in recovery plays a key role in improving
the overall recovery performance.  We show that compared to erasure codes,
CORE achieves recovery throughput gains with up to $3.4\times$ for single
failures and up to $2.3\times$ for concurrent failures.  Our experimental
results conform to our theoretical findings.  We also evaluate the runtime
performance of MapReduce jobs under node failures.  We show that CORE can
reduce the runtime of a MapReduce job in both single and concurrent failures 
when compared to erasure codes.  Furthermore, our prototype maintains the
performance of striping replicas into encoded data, an operation that is
included in original HDFS-RAID, when regenerating codes are used. 
\end{itemize}

The rest of the paper proceeds as follows. 
Section~\ref{sec:model} first formulates our system model. 
Section~\ref{sec:motivating} motivates how CORE reduces bandwidth of
conventional recovery. 
Section~\ref{sec:design} describes the design of CORE and presents our
theoretical and analysis findings.
Section~\ref{sec:implementation} describes the implementation details of CORE.
Section~\ref{sec:experiment} presents experimental results.
Section~\ref{sec:related} reviews related work.
Section~\ref{sec:discussion} discusses several open issues of CORE, and finally,
Section~\ref{sec:conclusion} concludes this paper.

\section{System Model}
\label{sec:model}

We formulate the recovery problem in a distributed storage system. We also
provide an overview of regenerating codes, and show how they can improve the
recovery performance.
%Our work covers a general number of failures, including both single and
%concurrent failures.  
%We use a relayer model, in which a dedicated relayer node centrally
%coordinates the reconstruction process.  Based on this relayer model, we show
%how regenerating codes reduce the communication overhead of reconstruction. 

\subsection{Basics}
\label{subsec:basics}

We first define the terminologies and notation.  Table~\ref{tab:notation}
summarizes the major notation used in this paper. 
%some of which are  based on studies \cite{khan12,plank09}.  
We consider a distributed storage system composed of a collection of 
{\em nodes}, each of which refers to a physical storage device.  The storage
system contains $n$ nodes labeled by $N_0, N_1, \cdots, N_{n-1}$, in which $k$
nodes (called {\em data nodes}) store the original (uncoded) data and the
remaining $n-k$ nodes (called {\em parity nodes}) store parity (coded) data.
The coding structure is {\em systematic}, meaning that the original data is
kept in storage. 

\begin{figure}[t]
\centering
\includegraphics[width=4in]{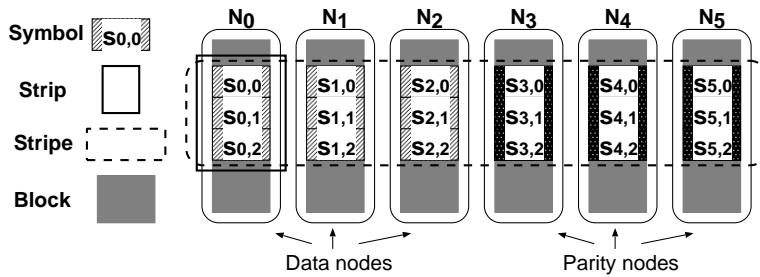}
\caption{Example of a distributed storage system, where $n=6$, $k=3$,
and $r=3$. We assume that nodes $N_0$, $N_1$, and $N_2$ are data nodes,  
while $N_3$, $N_4$, and $N_5$ are parity nodes. For load balancing, the
identities of data and parities nodes are rotated across different blocks.}
\label{fig:notation}
\end{figure}

Figure~\ref{fig:notation} shows an example of a distributed storage
system, which is also consistent with the erasure-coded design of HDFS-RAID
\cite{hdfsraid}.  Each node stores a number of {\em blocks}. A block is the
basic unit of read/write operations in a storage system.  It is called
a data block if it holds original data, or a parity block if it holds parity
data.  To store data/parity information, each block is partitioned into
fixed-size {\em strips}, each of which contains $r$ {\em symbols}.  A symbol
is the basic unit of encoding/decoding operations.  A {\em stripe} is a
collection of strips on $k$ data nodes and the corresponding encoded strips on
$n-k$ parity nodes.  A data (parity) block contains all strips of data
(parity) symbols.  For load balancing reasons the identities of the
data/parity nodes are rotated so that the data and parity blocks are evenly
distributed across nodes \cite{khan12,plank09}. 

\begin{table}[t]
\centering
\caption{Major notation used in this paper.}
\begin{small}
\begin{tabular}{|p{0.3in}|p{5.7in}|}
\hline
$n$ & number of nodes\\
\hline
$N_i$ & the $i$-th node ($0\le i\le n-1$)\\
\hline
$k$ & number of data nodes\\
\hline
$r$ & number of symbols per strip\\
\hline
$t$ & number of concurrent failures ($1\le t\le n-k$)\\
\hline
$M$ & size of original data stored in a stripe\\
\hline
$s_{i,j}$ & the $j$-th stored symbol in a stripe of node $N_i$ ($0\le i\le
		n-1$, $0\le j\le r$)\\
\hline
$e_{i,i'}$ & encoded symbol from surviving node $N_i$ used to recover
lost data of failed node $N_{i'}$ ($0\le i, i'\le n-1$)\\
\hline
\end{tabular}
\end{small}
\label{tab:notation}
\end{table}

Each stripe is independently encoded.  Our discussion thus focuses on a single
stripe and our recovery scheme will operate on a per-stripe basis.  
Let $M$ be the total amount of original uncoded data stored in a stripe.  
Let $s_{i,j}$ be a stored symbol of node $N_i$ at offset $j$ in a stripe, where
$i=0,1,\cdots,n-1$ and $j=0,1,\cdots r-1$.  Each stripe contains $nr$
stored symbols, which can be formed by multiplying an $nr \times kr$ 
{\em generator matrix} by a vector of $kr$ original data symbols based on the
{\em Galois field} arithmetic, whose implementation details can be found in
the prior study \cite{greenan08}.  In this work, we focus on the arithmetic
operations over the Galois field GF($2^8$).  Note that our recovery scheme
applies to the failures of both data and parity nodes.  It treats each stored
symbol $s_{i,j}$ the same way regardless of whether it is a data or parity
symbol. 

For data availability, we have the storage system employ an $(n,k)$ code that
is {\em maximum distance separable (MDS)}, meaning that the stored data of any
$k$ out of the $n$ nodes can be used to reconstruct the original data.  That
is, an $(n,k)$ MDS-coded storage system can tolerate any $n-k$ out of $n$
concurrent failures.  MDS codes also ensure optimal storage efficiency, such
that each node stores $\frac{M}{k}$ units of data per stripe.  Reed-Solomon
(RS) codes \cite{reed60} are a classical example of MDS codes.  RS codes can
be implemented with strip size $r = 1$ to minimize the generator matrix size. 

\subsection{Recovery}
\label{subsec:model_recovery}

Our recovery addresses two types of node failures. The first type is the
recovery from permanent failures (e.g., due to crashes) where data is
permanently lost.  In this case, we reconstruct the lost data of the failed
nodes on new nodes to minimize the window of vulnerability.  Another type is
degraded reads to the temporarily unavailable data during transient failures 
(e.g., due to system reboots or upgrades) or before the permanent failures are
restored.  The reads are degraded as the unavailable data needs to be
reconstructed from the available data of other surviving nodes.  In our
discussion, we use ``lost data'' to refer to both permanently lost data and
temporarily unavailable data.

We consider the scenario where the storage system activates recovery of lost
data when there are a number $t \ge 1$ of failed nodes.  Clearly, we require
$t \le n-k$, or the original data will be unrecoverable.  We call the set of
$t$ failed nodes the {\em failure pattern}.  The lost data will be
reconstructed by the data stored in other surviving nodes.

%The reconstruction is said to be {\em eager} if it is activated when there
%exists a single node failure (i.e., $t=1$), or {\em lazy} if it is activated
%only when there are multiple failures (i.e., $t>1$). Lazy reconstruction is
%first studied in \cite{bhagwan04}, and it uses erasure codes. 

%Now let us consider multi-node failure recovery. Let $t$ be the number of
%failed nodes. Let $F$ be the set containing the $t$ failed node, we call $F$
%a failure pattern.  Thus, there will be $t$ newcomers when recovering the
%failures. 
%One of the newcomers will download data from surviving nodes,
%generate the lost data and forward the data to other newcomers. This 
%newcomer is called \emph{relayer}. In our relayer model, we will 
%propose a repair scheme which connects to $d'=d+1-t$ surving nodes to
%recover a $t$-node failure.

\begin{figure}
\centering
\includegraphics[width=3in]{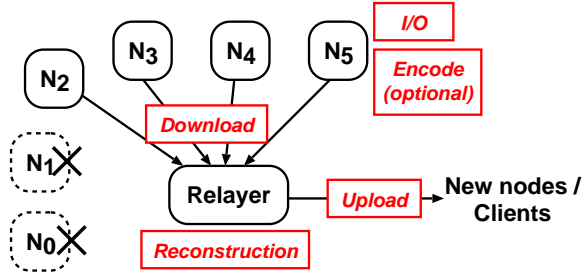}
\caption{Recovering nodes $N_0$ and $N_1$ using the relayer model.}
\label{fig:relayer_model}
\end{figure}

Our recovery builds on the {\em relayer} model, in which a relayer daemon
coordinates the recovery operation.  Figure~\ref{fig:relayer_model} depicts
the relayer model.  During recovery, each surviving node performs two steps:
(i) {\em I/O}: it reads its stored data, and (ii) {\em encode} (for
regenerating codes only): it combines the stored data into some linear
combinations.  The relayer daemon performs three steps: (i) {\em download}: it
downloads the data from some other surviving nodes, (ii) {\em reconstruction}:
it reconstructs the lost data, and (iii) {\em upload}: it uploads the
reconstructed data to the new nodes (for recovery from permanent failures) or
to the client who requests the data (for degraded reads).  We assume that the
relayer is reliable during the recovery process.  

We argue that the relayer model can be easily fit into practical distributed
storage systems.  In the case of recovering permanent failures, we can deploy
the relayer daemon in different ways, such as in one of the new storage nodes
that reconstructs all lost data, in every storage node that reconstructs a
subset of lost data, or in separate servers that run outside the storage
system.  In the case of degraded reads, we can deploy the relayer daemon in
each storage client.  We note that this relayer model is also used in prior
studies in the contexts of peer-to-peer storage \cite{bhagwan04}, data center
storage \cite{huang12}, and proxy-based cloud storage \cite{hu12}.    In
Section~\ref{sec:implementation}, we elaborate how the relayer model can be
integrated into a distributed storage system. 

%We point out that concurrent recovery has a side benefit of delaying immediate
%recoveries.  That is, we can perform recovery only when the number of failures
%exceeds a tolerable limit.  This avoids unnecessary recoveries should a
%failure be transient and the data be available shortly (e.g., after rebooting
%a failed node).  This aligns with the concept of lazy repair in
%\cite{bhagwan04}.  

To improve the recovery performance of a distributed storage system with
limited network bandwidth, it is important to minimize the amount of data
transferred over the network.  If the number of failed nodes is small, the
amount of data being downloaded from the surviving nodes is larger than the
amount of reconstructed data being uploaded to new nodes (or clients).  If we
pipeline the download and upload steps (see Section~\ref{subsec:integration}),
then the download step becomes the bottleneck.
Thus, we focus on optimizing the download step in recovery.
Formally, we define the {\em recovery bandwidth} as the total amount of data
being downloaded per stripe from the surviving nodes to the relayer during
recovery.   Our goal is to minimize the recovery bandwidth. 

%If we pipeline the download and upload steps (see
%Section~\ref{subsec:integration}), then the download step will become %the
%bottleneck.  
%Minimizing the recovery bandwidth is necessary for a distributed storage
%system with limited network bandwidth.
%Here, we assume that there
%is limited data transmission bandwidth in the %distributed storage system we
%consider. Thus, our goal is to minimize the %reconstruction bandwidth.  

\subsection{Regenerating Codes}
\label{subsec:regenerating}

When an erasure-coded system sees failures, 
{\em conventional recovery} is used, meaning that the relayer downloads data
from any $k$ surviving nodes to first reconstruct all original data and then
return the lost data.  The amount of data being downloaded is equal to the
amount of original data being stored (i.e., $M$ per stripe).  Note that some
proposals allow less data to be read for some erasure codes under specific
conditions (see Section~\ref{sec:related}).  However, conventional recovery
applies to {\em any MDS erasure code and any number of failures no more than
$n-k$}.  In this paper, when we refer to erasure codes, we assume that
conventional recovery is used. 

We consider a special class of codes called {\em regenerating codes}
\cite{dimakis10} that enables the relayer to transfer less than the amount of
original data being stored.  Regenerating codes build on network coding
\cite{ahlswede00}, in which during recovery, surviving nodes send encoded
symbols that are computed by the linear combinations of their stored symbols,
and then the encoded symbols are used to reconstruct the lost data. 
It is shown that regenerating codes lie on an optimal tradeoff curve between
storage cost and recovery bandwidth \cite{dimakis10}.  There are two extreme
points: {\em minimum storage regenerating (MSR)} codes, in which each node
stores the minimum amount of data on the tradeoff curve, and {\em minimum
bandwidth regenerating (MBR)} codes, in which the bandwidth is minimized.
Note that MSR codes have the same optimal storage efficiency as MDS erasure
codes such as RS codes, while MBR codes minimizes bandwidth at the expense of
higher storage overhead.  In this work, we focus on MSR codes. 

Existing optimal MSR codes are designed for recovering a single failure, as
described below.  First, the strip size has $r = n-k$ symbols to achieve the
minimum possible bandwidth.  During recovery, the relayer downloads one 
{\em encoded} symbol from each of the $n-1$ surviving 
nodes\footnote{There are MSR code constructions (e.g., \cite{suh11,rashmi11b})
that can download encoded symbols from less than $n-1$ surviving nodes at the
expense of higher recovery bandwidth.  In this work, we only focus on the case
where $n-1$ surviving nodes are connected.}.
Let $e_{i,i'}$ be the encoded
symbol downloaded from node $N_i$ and used to reconstruct data for the failed
node $N_{i'}$.  Each encoded symbol $e_{i,i'}$ is a function of the symbols
$s_{i,0}, s_{i,1}, \cdots, s_{i,r-1}$ stored in the surviving node $N_i$, and
has the same size as each stored symbol.  Using the encoded symbols, the
relayer reconstructs the lost symbols of the failed node $N_{i'}$.  MSR codes
achieve the minimum recovery bandwidth (denoted by $\gamma_{MSR}$) for single
failure recovery given by \cite{dimakis10}:
\begin{equation}
\gamma_{MSR} = \frac{M(n-1)}{k(n-k)}.
\label{eqn:msr}
\end{equation}

However, existing studies on regenerating codes are limited in different
aspects, which we further discuss in Section~\ref{sec:related}. To summarize,
most recovery approaches focus on single failures.  If more than one node
fails, the optimal MSR code constructions cannot achieve the saving shown in
Equation~(\ref{eqn:msr}) by connecting to $n-1$ surviving nodes.  To recover
concurrent failures, a straightforward approach is to resort to conventional
recovery and download the size of original data from any $k$ surviving nodes.
This paper explores if we can achieve recovery bandwidth saving for concurrent
failures as well.

\section{Motivating Example}
\label{sec:motivating}

Before we describe the design of CORE, we first motivate via an example how
CORE reduces the recovery bandwidth over conventional recovery for concurrent
failures.  The design details of CORE will be refined in
Section~\ref{sec:design}.

%Recall that the recovery operates on a per-stripe basis.  For each stripe, we
%have $k=3$ data nodes (e.g., $N_0$, $N_1$, $N_2$) store original data
%symbols, and the other $n-k = 3$ parity nodes (e.g., $N_3$, $N_4$, $N_5$)
%store parity symbols.
We consider an MDS code with $n=6$ and $k=3$.  Suppose that we store a data
object of size $M$ that corresponds to a stripe of original data symbols.  For
erasure codes, the strip size is $r = 1$ symbol, and the symbol size is
$\frac{M}{3}$.  For regenerating codes, the strip size is set to $r = n - k =
3$ symbols, and hence the symbol size is $\frac{M}{9}$.  Suppose now both
nodes $N_0$ and $N_1$ fail. Our goal is to reconstruct their lost data. 

\begin{figure}[t]
\centering
\begin{tabular}{cc}
%\vspace{0.1in}
%\hspace{-0.1in}
\hspace{-10pt}
\includegraphics[width=0.4\linewidth]{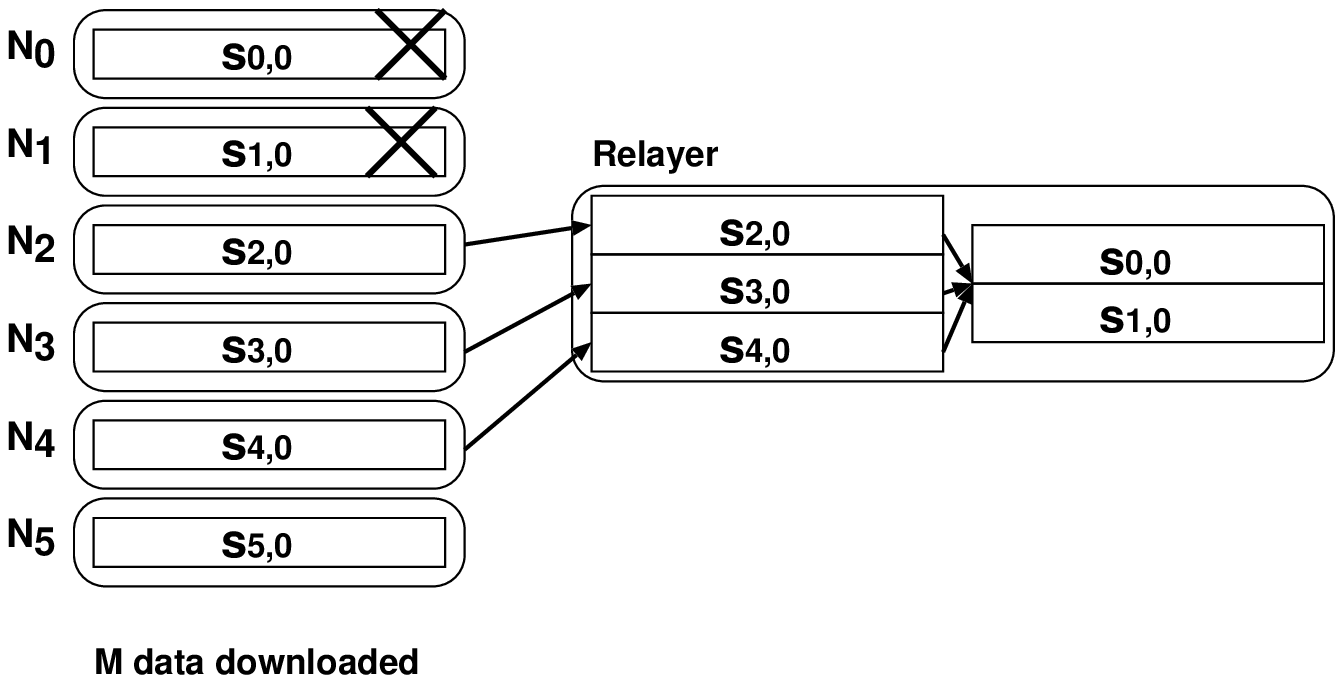} &
\hspace{-12pt}
\includegraphics[width=0.4\linewidth]{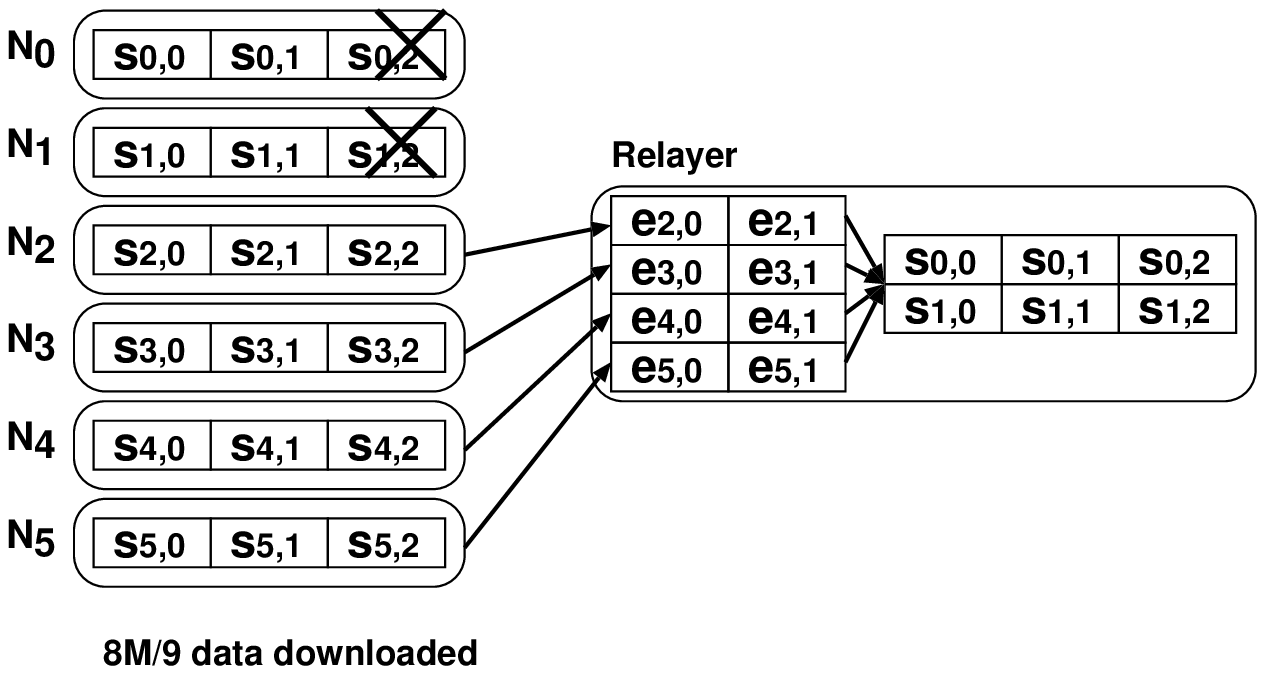} \\
%\hspace{-0.1in}
\mbox{\small (a) Conventional recovery (based on erasure codes)} &
\mbox{\small (b) CORE (see Section~\ref{sec:design})}
\end{tabular}
\caption{Comparisons of conventional recovery and CORE.}
\label{fig:comparisons}
\end{figure}

We first consider conventional recovery based on erasure codes, whose 
idea is to first reconstruct all original data.  Thus, the relayer downloads
the size of original data from any $k=3$ nodes (e.g., $N_2, N_3, N_4$).
Figure~\ref{fig:comparisons}(a) shows the conventional recovery, in which the
relayer reconstructs all three original symbols to regenerate the data for the
two failed nodes simultaneously. Thus, the total amount of data downloaded is
$M$. 
	
%shows the eager recovery.  Each time the relayer reads a size $M$ of
%data, so the total amount of data read is $2M$ for recovering both $N_0$ and
%$N_1$.  

We now discuss how CORE applies concurrent failure recovery.  Here, we
consider the baseline approach of CORE (see Section~\ref{subsec:recovery}).
Figure~\ref{fig:comparisons}(b) shows the main idea, in which the relayer now
downloads two encoded symbols $e_{i,0}$ and $e_{i,1}$ from each of the four
surviving nodes $N_i$ ($i=2,3,4,5$), such that CORE form a system of equations
in $e_{i,0}$'s and $e_{i,1}$'s to reconstruct the lost data of nodes $N_0$ and
$N_1$.  The total amount of data downloaded is $\frac{8M}{9}$, which is one
symbol size less than that of conventional recovery.  We point out that the
bandwidth saving of CORE can be even higher for some parameters.  In general,
when the baseline approach of CORE applies concurrent failure recovery for $t$
failures, the relayer downloads $t$ encoded symbols from each of the $n-t$
surviving nodes.  We elaborate the details in the next section.

%We now show that our concurrent recovery also provides bandwidth saving
%over immediate recovery.  Suppose that $N_0$ first fails, and then $N_1$
%fails.  If we reconstruct each individual failure using regenerating codes,
%then the total amount of data downloaded is $\frac{10M}{9}$ (by substituting
%$d=n-1$ and $k=3$ into Equation~(\ref{eqn:msr})), which is even more than the
%amount $M$ used by the conventional recovery. 

%Figure~\ref{fig:comparisons}(d) shows the eager recovery.  Each time the relayer reads one symbol from each of the five surviving nodes, so the total amount of data read is $\frac{10M}{9}$ for recovering both $N_0$ and $N_1$, and is less than that of the eager recovery for erasure codes.

%Section~\ref{sec:design} elaborates the design of CORE and shows its
%effectiveness. 

\section{Design of CORE}
\label{sec:design}

CORE builds on existing MSR code constructions that are designed for single
failure recovery with parameters $(n,k)$.  CORE has two major design goals.
First, CORE preserves existing code constructions and stored data.  That
is, we still have data striped and stored with existing MSR code
constructions, while CORE sits as a layer atop existing MSR code constructions
and enables efficient recovery for both single and concurrent failures.  The
optimal storage efficiency of MSR codes is still preserved.  Second, CORE aims
to minimize recovery bandwidth for a variable number $t\le n-k$ of concurrent
failures, without requiring $t$ to be fixed before a code is constructed and
the data is stored. 

In this section, we first describe the baseline approach of CORE, in which we
extend the existing optimal solution of single failure recovery to support
concurrent failure recovery (Section~\ref{subsec:recovery}).  We note that the
baseline approach of CORE is not applicable in a small proportion of failure
patterns, so we propose a simple extension that still provides bandwidth
reduction for such cases (Section~\ref{subsec:counter}). We present
theoretical results showing that CORE can reach the optimal point for a
majority of failure patterns (Section~\ref{subsec:theoretical}).  Finally, we
analyze the recovery bandwidth saving (Section~\ref{subsec:analysis}) and 
reliability (Section~\ref{subsec:mttf}) of CORE. 

\subsection{Baseline Approach of CORE}
\label{subsec:recovery}

%However, enabling existing MSR codes for reconstructing concurrent failures is
%challenging.  Recall that MSR codes are optimized for single failure
%recovery and the recovery bandwidth is minimized at $d=n-1$ (see
%Equation~(\ref{eqn:msr})). This implies that the relayer must connect to $n-1$
%surviving nodes for recovery.  However, connecting to $n-1$ surviving
%nodes becomes infeasible when there are multiple failures (i.e., $t > 1$).
%Some existing designs of MSR codes (e.g., \cite{rashmi11b,suh11}) allow some
%smaller values of $d$ while satisfying Equation~(\ref{eqn:msr}) (note that the
%achievable minimum bandwidth is higher than that of $d = n-1$).  However, it
%requires $d$ be fixed {\em in advance} before the MSR codes are constructed.
%Since the number $t$ of failed nodes is a variable in practice, it is
%non-trivial to pick the best $d$. 

%We support {\em any} MSR codes (e.g., \cite{hu12,rashmi11b,suh11}) to achieve
%the minimum possible recovery bandwidth shown in
%Equation~(\ref{eqn:msr}).  We also allow a variable number $t \le n-k$ of
%concurrent failures to be reconstructed. 

We first provide the background of existing MSR code constructions on which
CORE is developed.   We then define the building blocks of CORE, and explain
how CORE uses these building blocks to support concurrent failure recovery. 

{\bf Background.} CORE can build on existing optimal MSR code constructions
including Interference Alignment (IA) codes \cite{suh11} and Product-Matrix
(PM) codes \cite{rashmi11b}.  We here provide a high-level overview of how IA
codes work, while PM codes have a similar idea.  IA codes extend the idea of
aligning interference signals in wireless communication into failure recovery
in distributed storage systems.  Recall that each stripe in regenerating codes
contains $k(n-k)$ original data symbols (see Section~\ref{subsec:basics}).
Each stored symbol is a linear combination of the $k(n-k)$ original data
symbols.  Suppose that a data node fails (the similar idea also applies for
parity nodes).  The $n-1$ surviving nodes compute the $n-1$ encoded symbols
(denoted by ${\mathbf y} = (y_1, \cdots, y_{n-1})^T$).  The relayer downloads
the $n-1$ encoded symbols and reconstructs the $n-k$ lost data symbols
(denoted by ${\mathbf x_1} = (x_1, \cdots, x_{n-k})^T$) of the failed node.
There are other $(k-1)(n-k)$ data symbols (denoted by ${\mathbf x_2} =
(x_{(n-k)+1}, \cdots, x_{k(n-k)})^T$) that do not need to be regenerated and
can be viewed as interference signals.   We can express ${\mathbf y}$ as a
system of equations in ${\mathbf x_1}$ and ${\mathbf x_2}$ as:
\begin{displaymath}
\left(\!
\begin{array}{c}
{\mathbf A} \ \big\vert \ {\mathbf B}\\
\end{array}
\!\right) 
\left(
\begin{array}{c}
{\mathbf x_1}\\
{\mathbf x_2}
\end{array}
\right) 
= 
{\mathbf y}, 
\end{displaymath}
for some coefficient matrices ${\mathbf A}$ and ${\mathbf B}$ of sizes
$(n-1)\times(n-k)$ and $(n-1)\times(k-1)(n-k)$, respectively.  By elementary
row operations, we can transform the system of equations into:
\begin{displaymath}
\left(\!
\begin{array}{ccc}
{\mathbf A'}\!\!&\!\!\big\vert\!\!&\!\!{\mathbf 0}\\
{\mathbf 0}\!\!&\!\!\big\vert\!\!&\!\!{\mathbf B'}\\
\end{array}
\!\right) 
\left(
\begin{array}{c}
{\mathbf x_1}\\
{\mathbf x_2}
\end{array}
\right) 
= 
{\mathbf y'}, 
\end{displaymath}
for transformed vector ${\mathbf y'}$ and transformed matrices ${\mathbf A'}$
and ${\mathbf B'}$ of sizes $(n-k)\times(n-k)$ and
$(k-1)\times(k-1)(n-k)$, respectively.
Note that IA codes ensure that there
exists some transformation that makes ${\mathbf A'}$ an invertible matrix, so
that ${\mathbf x_1}$ (i.e., the lost symbols) can be uniquely solved.  

IA codes design the generator matrix that satisfies the above properties.  PM
codes have a similar idea using a different generator matrix design.  We refer
readers to \cite{suh11,rashmi11b} for their mathematical details on the
generator matrix design. 

Note that both IA and PM codes have parameter constraints.
IA codes require $n\geq 2k$, and PM codes require $n\geq 2k-1$.  In this work,
we mainly focus on the double redundancy $n=2k$, which is also considered in
state-of-the-art distributed storage systems such as OceanStore
\cite{kubiatowicz00} and CFS \cite{dabek01}. While the redundancy overhead is
higher than traditional RAID-5 and RAID-6 codes for large ($n$,$k$), it
remains less than traditional 3-way replication used in production storage
systems such as GFS \cite{ghemawat03} and HDFS \cite{shvachko10}. 

{\bf Building blocks.} Our observation is that any optimal MSR code
construction can be defined by two functions. Let {\sf Enc}$_{i,i'}$ be the
encoding function that is called by node $N_i$ to generate an encoded symbol
$e_{i,i'}$
for the failed node $N_{i'}$ using the $r = n-k$ stored symbols in node $N_i$ as
inputs; let {\sf Rec}$_{i'}$ be the reconstruction function that returns the
set of $n-k$ stored symbols of a failed node $N_i'$ using the encoded symbols
from the other $n-1$ surviving nodes as inputs.  Both {\sf Enc} and {\sf Rec}
define the operations of linear combinations of the stored symbols
$s_{i,j}$'s, depending on the specific code construction.  From the above
discussion, {\sf Enc} is to construct the encoded symbols $\mathbf{y}$, while
{\sf Rec} is to reconstruct the lost symbols $\mathbf{x_1}$. 

CORE works for {\em any} construction of optimal MSR codes, as long as the
functions {\sf Enc} and {\sf Rec} are well-defined.  The two functions 
{\sf Enc} and {\sf Rec} form the building blocks of CORE. 
	
{\bf Main idea of the baseline approach.} We consider two types of encoded
symbols to be downloaded for recovery: {\em real symbols} and {\em virtual
symbols}.  To recover each of the $t$ failed nodes, the relayer still operates
as if it connects to $n-1$ nodes, but this time it represents the symbols to
be downloaded from the failed nodes as virtual symbols, while still
downloading the symbols from the remaining $n-t$ surviving nodes as real
symbols.  Now, using {\sf Enc} and {\sf Rec}, we reconstruct each virtual
symbol as a function of the downloaded real symbols.  Finally, using the
downloaded real symbols and the reconstructed virtual symbols, we can
reconstruct the lost stored symbols in the failed nodes. 

\begin{figure}
\centering
\includegraphics[width=3in]{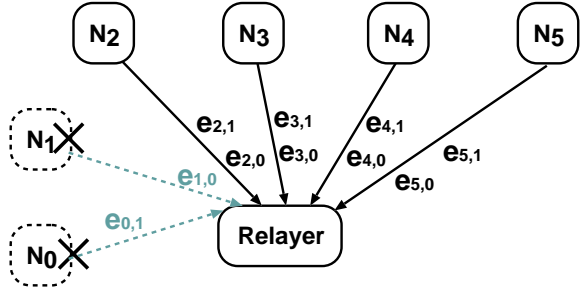}
\caption{An example of how the relayer downloads real and virtual symbols for
a (6,3) code when there are two failed nodes $N_0$ and $N_1$.  Here, $e_{1,0}$
and $e_{0,1}$ are the virtual symbols.}
\label{fig:virtual_symbols}
\end{figure}

{\bf Example.}
We depict our idea using Figure~\ref{fig:virtual_symbols}, which shows a
(6,3) code and has failures $N_0$ and $N_1$.  The two encoded symbols
$e_{1,0}$ and $e_{0,1}$ are virtual symbols, and the rest are real symbols.
We can express $e_{1,0}$ and $e_{0,1}$ based on the functions {\sf Enc} and
{\sf Rec} for single failure recovery as: 
\begin{eqnarray*}
e_{1,0} & = & \textrm{{\sf Enc}}_{1,0}(s_{1,0}, s_{1,1}, s_{1,2})
		\ = \ \textrm{{\sf Enc}}_{1,0}(\textrm{{\sf Rec}}_1(e_{0,1}, e_{2,1},
			e_{3,1}, e_{4,1}, e_{5,1}))\\
e_{0,1} & = & \textrm{{\sf Enc}}_{0,1}(s_{0,0}, s_{0,1}, s_{0,2})
		\ = \ \textrm{{\sf Enc}}_{0,1}(\textrm{{\sf Rec}}_0(e_{1,0}, e_{2,0},
			e_{3,0}, e_{4,0}, e_{5,0}))
\end{eqnarray*}
The encoded symbol $e_{1,0}$ is computed by encoding the stored symbols
$s_{1,0}$, $s_{1,1}$, and $s_{1,2}$, all of which can be reconstructed from
other encoded symbols $e_{0,1}$, $e_{2,1}$, $e_{3,1}$, $e_{4,1}$, and
$e_{5,1}$ based on single failure recovery.  Thus, $e_{1,0}$ can be expressed
as a function of encoded symbols.  The encoded symbol $e_{0,1}$ is expressed
in a similar way.  Now, we have two equations with two unknowns $e_{1,0}$ and
$e_{0,1}$.  If these two equations are linearly independent, we can solve for
$e_{1,0}$ and $e_{0,1}$. 
Then we can apply {\sf Rec}$_{0}$ and {\sf Rec}$_{1}$ 
to reconstruct the lost stored symbols of $N_{0}$ and $N_{1}$.  
In general, to recover $t$ failed nodes,
we have a total of $t(t-1)$ virtual symbols.  We can compose $t(t-1)$
equations based on the above idea.  If these $t(t-1)$ equations are linearly
independent, we can solve for the virtual symbols.  A subtle issue is that the
system of equations may be unsolvable.  We explain how we generalize our
baseline approach for such an issue in the next subsection.

\subsection{Recovering Any Failure Pattern}
\label{subsec:counter}

We seek to express the virtual symbols as a function of real symbols by
solving a system of equations.  However, we note that for some failure
patterns (i.e., the set of failed nodes), the system of equations cannot
return a unique solution.  A failure pattern is said to be {\em good} if we
can uniquely express the virtual symbols as a function of the real symbols, or 
{\em bad} otherwise.  Our goal is to reduce the recovery bandwidth even for
bad failure patterns. 

We first evaluate the likelihood of having bad failure patterns for different
choices of parameters. Given an $(n,k)$ code and $t$ failures, there are
$n\choose t$ possible failure patterns.  We enumerate all such possible
failure patterns and check if each of them is bad.  In practice, each stripe
has a limited number of nodes (i.e., $n$ will not be too large)
\cite{khan12,plank09}, so we can feasibly enumerate all possible failure
patterns and identify the bad ones in advance.  We conduct our enumeration
for both IA and PM codes. 

\begin{figure}[t]
\centering
\begin{tabular}{cc}
    \includegraphics[width=3in]{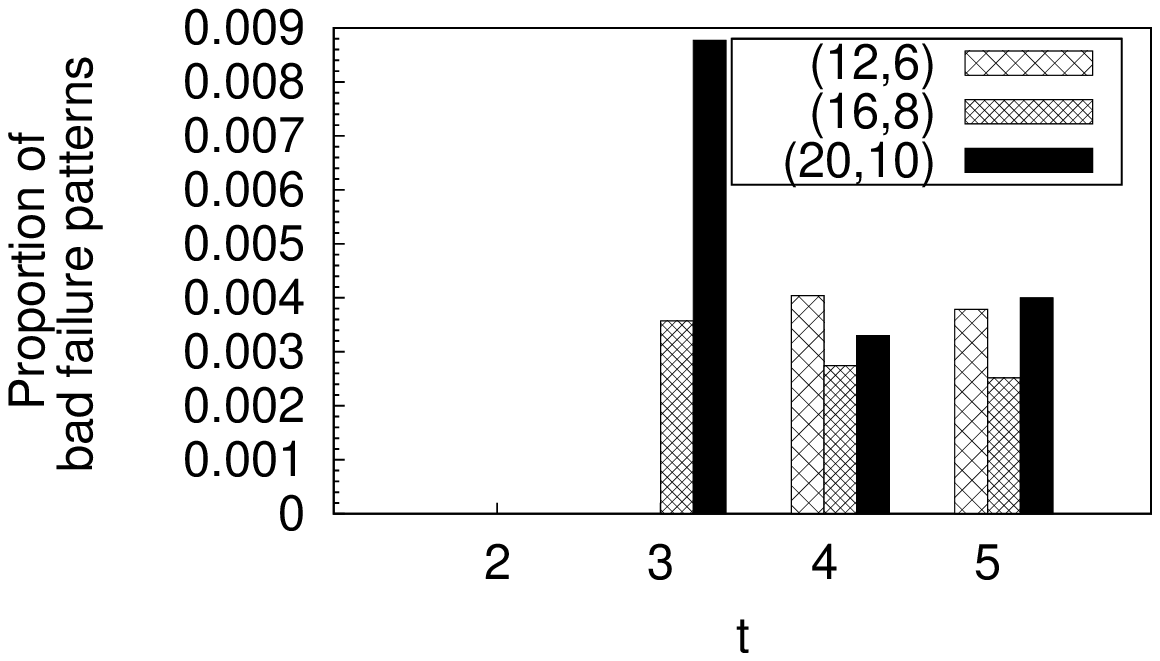} &
\includegraphics[width=3in]{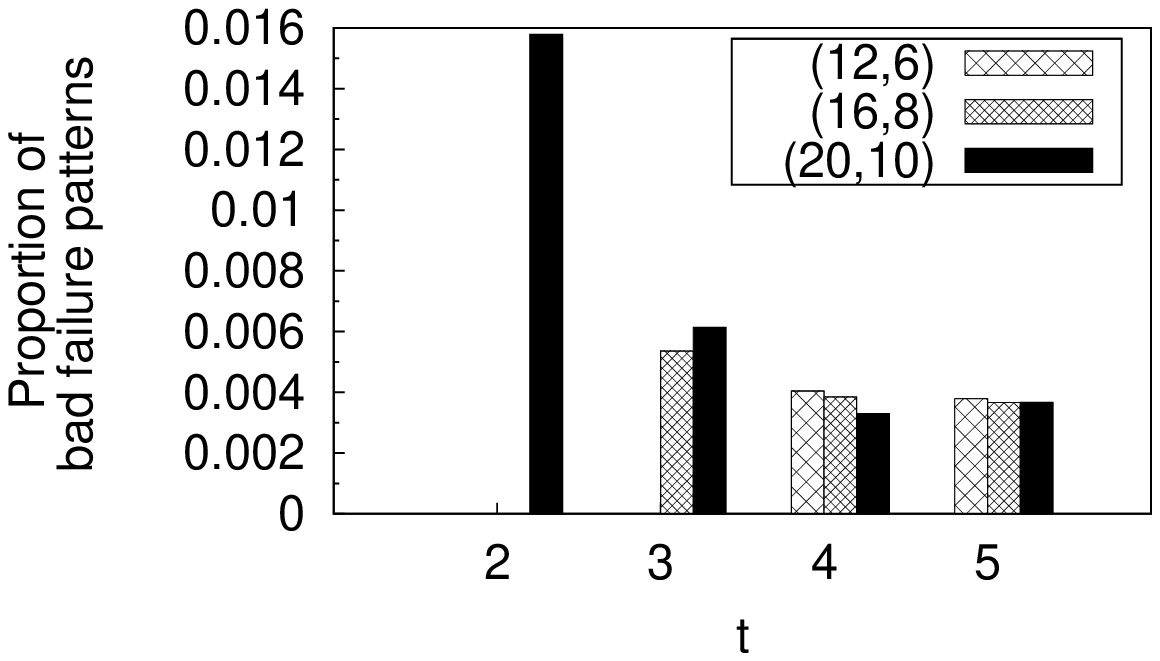} \\
\mbox{(a) IA codes} &
\mbox{(b) PM codes}
\end{tabular}
\caption{Proportions of bad failure patterns for different $(n,k)$ and $t$.}
\label{fig:bad}
\end{figure}

Figure~\ref{fig:bad} shows the proportions of bad failure patterns for
different combinations of $(n,k)$ and $t$.  We observe that among all
parameters we consider, bad failure patterns only account for a small
proportion, with at most 0.9\% and 1.6\% for IA and PM codes, respectively. 
Also, for some sets of parameters, we do not find any
bad failure patterns.  Nevertheless, we would like to reduce the recovery
bandwidth for such bad failure patterns even though they are rare. 

\begin{figure}[t]
\centering
\includegraphics[width=3in]{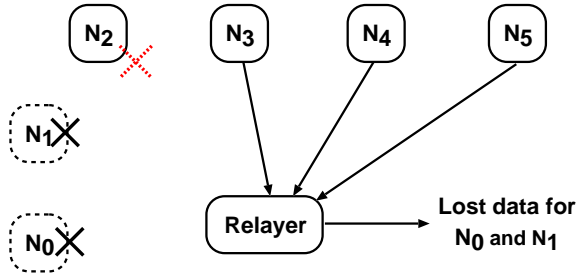}
\caption{An example of using a virtual failure pattern for a (6,3) code. If
the original failure pattern $\{N_0, N_1\}$ is bad, then we can instead
recover the virtual failure pattern $\{N_0, N_1, N_2\}$ and only download
encoded symbols from nodes $N_3, N_4, N_5$.}
\label{figVirtualFP}
\end{figure}

We now extend our baseline approach of CORE to deal with the bad failure
patterns, with an objective of reducing the recovery bandwidth over the
conventional recovery approach.  For a bad failure pattern $\SF$, we include
one additional surviving node and form a {\em virtual failure pattern} $\SF'$,
such that $\SF\subset\SF'$ and $|\SF'| = |\SF| + 1 = t+1$.  Then the relayer
downloads the data from the $n-t-1$ nodes outside $\SF'$ needed for
reconstructing the lost data of $\SF'$, although actually only the lost data
of $\SF$ needs to be reconstructed.  Figure~\ref{figVirtualFP} shows an
example of how we use a virtual failure pattern for recovery.  If $\SF'$
is still a bad failure pattern, then we include an additional surviving node
into $\SF'$, and repeat until a good failure pattern is found.  Note that the
size of $\SF'$ must be upper-bounded by $n-k$, as we can always connect to $k$
surviving nodes to reconstruct the original data due to the MDS code property. 

%Tradeoff of encode/decode performance and probability of bad failure
%patterns.
%In order to study the bad failure patterns, we tested a class of 
%$(2k-1,k)$ coding scheme with $k$ ranging from 3 to 8.
%We enumerate all the failure patterns to examine the probability of
%bad failure patterns. Recall we build the coding scheme in Galois field
%$GF(2^q)$. Table \ref{tbl_badprobability} shows the 
%probability of a bad failure pattern with different $q$ and $t$ value.
%By the result of this enumeration, we find that
%the larger the value of $q$, the less likely that a failure pattern is
%a bad failure pattern. When $q=8$, the bad failure patter probablity
%is less than $1\%$.

%By enumerating the failure patterns, we found that 
%when $q\geq 5$ for every failure pattern $F$, there always exists a 
%good failure pattern $F'$ satisfying $F\subset F'$ and $|F'|=|F|+1$.

\subsection{Theoretical Results}
\label{subsec:theoretical}

We present two theorems.  The first one shows the lower
bound of recovery bandwidth.  The second one shows that CORE achieves the
lower bound for good failure patterns.  The proofs are in Appendix. 

\begin{theorem}
Suppose that we recover $t$ failed nodes.  The lower bound of recovery
bandwidth is:
\[
\left\{ \ \
\begin{aligned}
& \frac{M t (n - t)}{k (n - k)} & \textrm{ where $t < k$,}\\
& M & \textrm{ where $t \ge k$.}
\end{aligned}
\right.
\vspace{-1em}
\]
$\hfill\Box$
\label{thm:thm1}
\end{theorem}

\begin{theorem}
CORE, which builds on MSR codes for single failure recovery, achieves the
lower bound in Theorem~\ref{thm:thm1} if we recover a good failure pattern. 
$\hfill\Box$
\label{thm:thm2}
\end{theorem}

Since most failure patterns are good (with at least 99.1\% and 98.4\% for IA
and PM codes, respectively), we conclude that CORE minimizes recovery
bandwidth for a majority of failure patterns. In the next subsection, we show
the actual bandwidth saving of CORE in both good and
failure patterns. 

\subsection{Analysis of Bandwidth Saving}
\label{subsec:analysis}

We now study the bandwidth saving of CORE over conventional recovery.  We
compute the bandwidth ratio, defined as the ratio of recovery bandwidth of
CORE to that of conventional recovery.  We vary $(n,k)$ and the number $t$ of
failed nodes to be recovered.

\begin{figure}[!t]
\centering
\begin{tabular}{cc}
\includegraphics[width=3in]{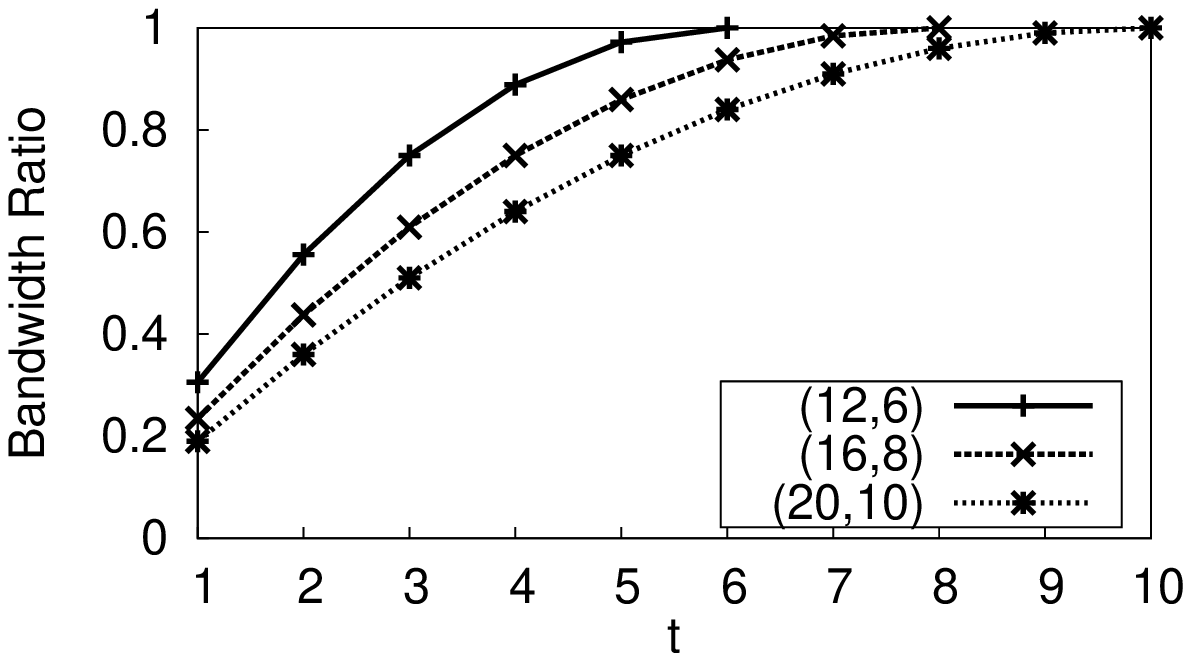} &
\includegraphics[width=3in]{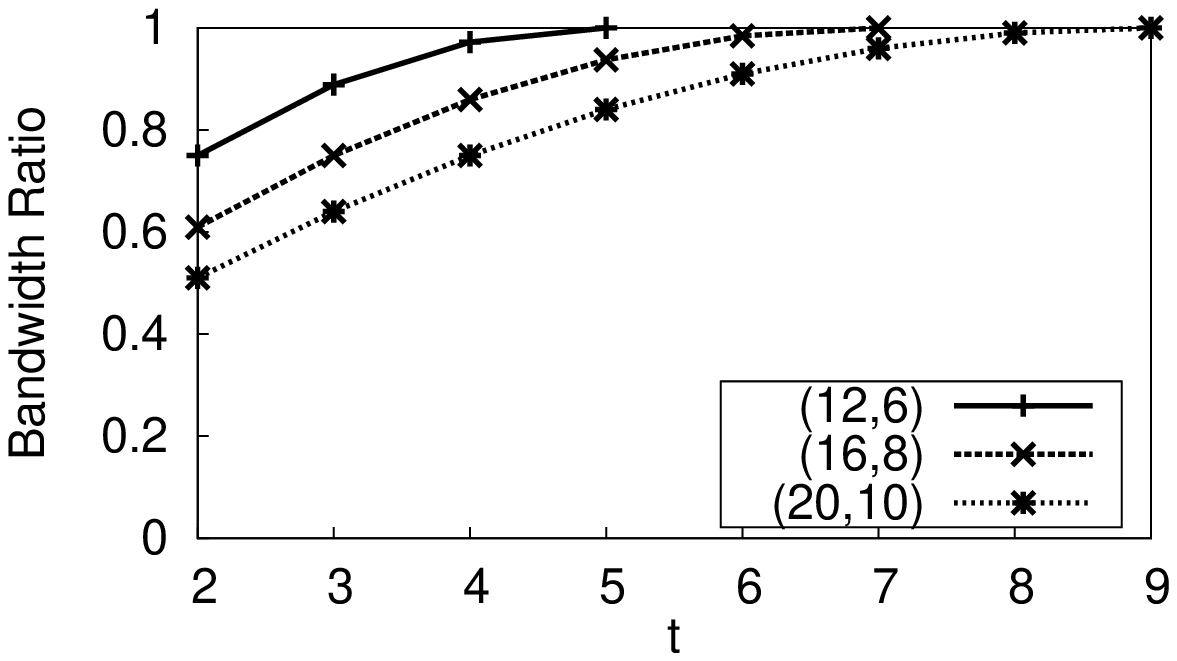}\\
\mbox{(a) Good failure patterns} &
\mbox{(b) Bad failure patterns}\\
\end{tabular}
\caption{Ratio of recovery bandwidth of CORE to that of conventional recovery.}
\label{fig:eval_bandwidth}
\end{figure}

We first consider good failure patterns.  For CORE, the recovery bandwidth
achieves the lower bound derived in Theorem~\ref{thm:thm1}, and we
can directly apply the theoretical results.  For conventional recovery, the
recovery bandwidth is the amount of original data being stored.
Figure~\ref{fig:eval_bandwidth}(a) shows the bandwidth ratio.  We observe
that CORE achieves bandwidth saving in both single and concurrent failures.
For single failures (i.e., $t=1$), CORE directly benefits from existing
regenerating codes, and saves the recovery bandwidth by 70-80\%.  For
concurrent failures (i.e., $t>1$), CORE also shows the bandwidth
saving, for example by 44-64\%, 25-49\%, and 11-36\% for $t=2$, $t=3$ and
$t=4$, respectively.  The bandwidth saving decreases as $t$ increases, since
more lost data needs to be reconstructed and we need to retrieve nearly the
amount of original data stored.  On the other hand, the bandwidth saving
increases with the values of $(n,k)$.  For example, the saving is 36-64\% in
(20,10) when $2\le t\le 4$. 

We now study how CORE performs for bad failure patterns.  
%We now show that this heuristic can still achieve reconstruction bandwidth
%saving compared to the conventional approach based on erasure codes.  
Recall from Section~\ref{subsec:counter} for each bad failure pattern $\SF$,
CORE forms a virtual failure pattern $\SF'$ that is a good failure
pattern. We compute the recovery bandwidth for $\SF'$ based on our theoretical
results in Section~\ref{subsec:theoretical}.
Figure~\ref{fig:eval_bandwidth}(b) shows the bandwidth ratio.  We find that in
all cases we consider, it suffices to add one surviving node into $\SF'$
(i.e., $|\SF'| = |\SF| + 1$) and obtain a good failure pattern.  Thus, the
recovery bandwidth of CORE for a bad $t$-failure pattern is always equivalent
to that for a good $(t+1)$-failure pattern.  From the figure, we still see
bandwidth saving of CORE over conventional recovery.  For example, the saving
is 25-49\% in (20,10) when $2\le t\le 4$. 
	
%We point out that concurrent reconstruction also outperforms immediate
%reconstruction in general.  Suppose that the $t$ failures show up one by one.
%In immediate reconstruction, we fix each single failure using MSR codes with
%$d=n-1$. Then the reconstruction bandwidth is $t\times \frac{M(n-1)}{k(n-k)}$
%(see Equation~(\ref{eqn:msr})). 

%Figure~\ref{fig:eval_bandwidth_good}(b) shows the ratio of the reconstruction
%bandwidth of concurrent reconstruction to that of immediate reconstruction,
%both of which use regenerating codes.  Our concurrent reconstruction
%shows the bandwidth saving in reconstructing multiple failures, for
%example by 6-17\% and 13-33\% when $t=2$ and $t=3$, respectively.  

%\begin{figure}[!t]
%\centering
%\includegraphics[width=0.95\linewidth]{figs/bad_bandwidth.eps}
%\vspace{-1em}
%\caption{The ratio of reconstruction bandwidth of regenerating codes to that
%of erasure codes for bad failure patterns.} 
%\label{fig:eval_bandwidth_bad}
%\end{figure}

%Although the benefits weaken as the number of failed nodes grows, the
%bandwidth can still be significantly reduced when the number of failed node
%is no larger than 3. 

\subsection{Analysis of Reliability}
\label{subsec:mttf}

We conduct reliability analysis on CORE and conventional recovery using the
Markov model.  Let $X$ be the random variable representing the time elapsed
until the data of a storage system becomes unrecoverable.  We define the
mean-time-to-failure (MTTF) as the expectation of $X$.  Prior studies have
also used the Markov model to analyze the reliability of systems with
replication (e.g., \cite{ford10}) and erasure codes (e.g.,
\cite{greenan09,huang12,sathiamoorthy13}).  Here, we focus on modeling the
reliability of CORE when concurrent failure recovery is used. 

\begin{figure}
\centering
%    \begin{tabular}{c}
\includegraphics[width=4in]{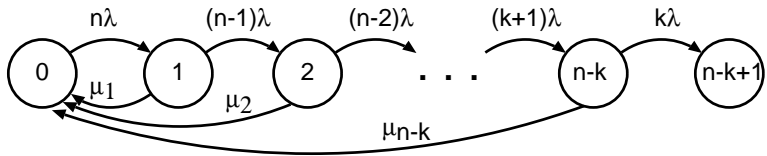}
%    \mbox{(a) Markov chain of relayer model}\\
%    \includegraphics[width=0.5\textwidth]{figs/mttf2.eps}\\
%    \mbox{(b) Markov chain of one-by-one recovey model}
%    \end{tabular}
\caption{Reliability model of $(n,k)$ codes.}
\label{fig:markovchain}
\end{figure}

Figure~\ref{fig:markovchain} shows the Markov model of $(n,k)$ codes.  Let
state~$t$, where $0\le t\le n-k$, denote that the storage system has $t$
failures, and state~$n-k+1$ denote that the storage system has more than $n-k$
failures and its data becomes unrecoverable.  To simplify the problem, we
assume that node failures occur independently and have constant rates as in
prior studies (e.g., \cite{ford10,greenan09,huang12,sathiamoorthy13}).  Let $\lambda$
denote the failure rate of a single node.  Thus, the transition rate from
state~$t$ (where $0\le t\le n-k$) to state $t+1$ is $(n-t)\lambda$.  In
concurrent recovery (assuming the relayer model in
Section~\ref{subsec:model_recovery} is used), 
every state~$t$ (where $1\le t\le n-k$) transitions to
state~0 at rate $\mu_t$, which depends on the recovery scheme being used. 
To compute $\mu_t$, let $B$ be the transfer rate of downloading data from
surviving nodes for recovery, and $S$ be the storage capacity of a single
storage node (i.e., the amount of original data is $kS$).  To recover $t$
failures, CORE downloads $\frac{t(n-t)}{k(n-k)}\times kS$ units of data in
most cases (see Theorems~\ref{thm:thm1} and \ref{thm:thm2})\footnote{Recall
that we assume $n=2k$ (see Section~\ref{subsec:recovery}), and hence $t < k =
n-k$ and we can apply Theorem~\ref{thm:thm1}.} and hence
$\mu_t=\frac{(n-k)B}{t(n-t)S}$; conventional recovery downloads $kS$ units of
data and hence $\mu_t=\frac{B}{kS}$.  Once the Markov model is constructed, we
can obtain the MTTF by calculating the expected time to reach the absorbing
state~$n-k+1$.  In the interest of space, we refer readers to \cite{greenan09}
for the detailed derivations of the MTTF. 

%The following two paragraphs shows how to calculate MTTF. 
%Let $\pi_i(t)$ represent the probability that at time $t$,
%the system is in state $i$.  We define vector  
%$\boldsymbol{\pi}=(\pi_0(t),\pi_1(t),...,\pi_{n-k+1}(t))$.  
%Since state $n-k+1$ represents data loss and is an absorbing
%state, thus $\pi_{n-k+1}(t)=Pr[X\le t]$, and $\pi_{n-k+1}'(t)$ is the 
%probability density function of random variable $X$.  Thus MTTF can be calculated by
%$$\int_0^{\infty}t\pi_{n-k+1}'(t)dt.$$
%
%To calculate MTTF, we need the rate matrix $Q$, which is derived from 
%the Markov model in Figure~\ref{fig:markovchain}(a). 
%Since $\boldsymbol{\pi}$ and $Q$ 
%satisfy that $\boldsymbol{\pi}'^T=\boldsymbol{\pi} Q$, we do 
%Laplace transform on the above equations.  
%Let $L_i(s)$ denote the Laplace transform of $\pi'_i(t)$.  
%After the Laplace transform, we get $n-k+2$ linear equations with 
%$L_i(s)(0\le i\le n-k+1)$ as unknowns, by solving the system of equations, 
%we can get the expressions of $L_i(s)$.  Since 
%$\int_0^{\infty}t\pi_{n-k+1}'(t)dt=-L_{n-k+1}'(s)|_{s=0}$, we can get the
%MTTF by calculating the value of $-L_{n-k+1}'(s)|_{s=0}$.
%
%Figure~\ref{fig:markovchain}(b) shows another recovery model, we name it
%one-by-one model.  Instead of recovering all failed nodes in one
%shot, like relayer model, one-by-one model recovers one failed node
%a time until all failed nodes are recovered.  One-by-one model is
%deployed by some systems (e.g., \cite{huang12,sathiamoorthy13}).

\begin{figure}
    \begin{tabular}{cc}
        \includegraphics[width=0.45\textwidth]{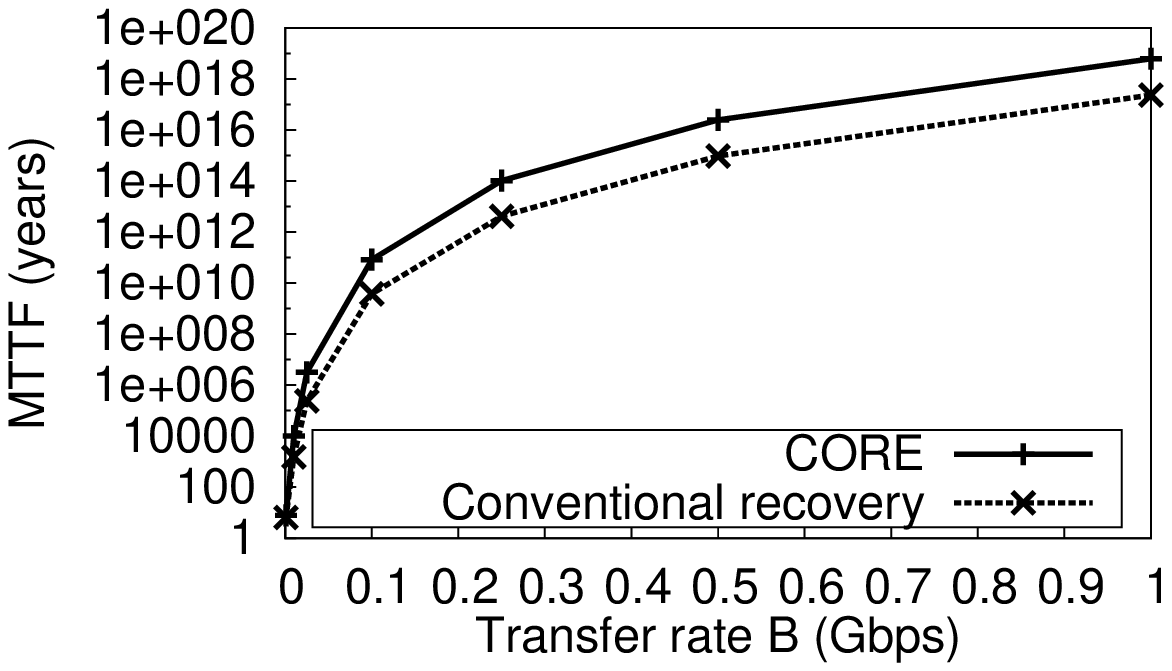} &
        \includegraphics[width=0.45\textwidth]{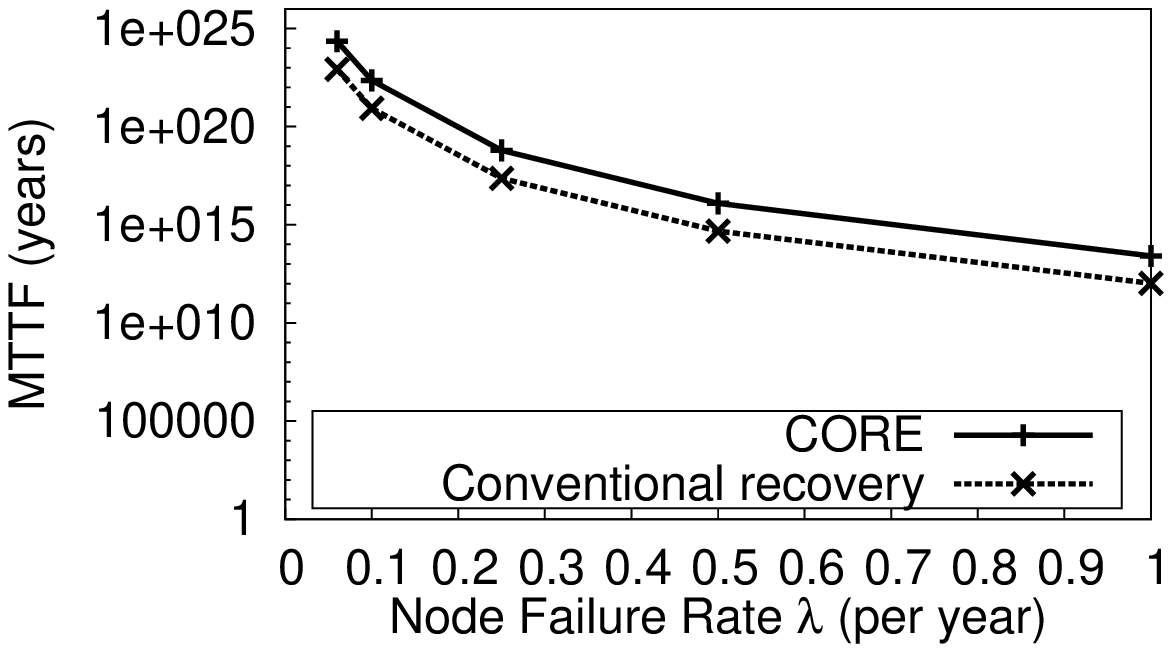} \\
        \mbox{(a) MTTF vs. transfer rate $B$} &
        \mbox{(b) MTTF vs. node failure rate $\lambda$}\\
    \end{tabular}
    \caption{Comparison of MTTF of CORE and conventional recovery.}
    \label{fig:mttfanalyze}
\end{figure}

We use $(n,k) = (16,8)$ as an example to compare the MTTFs of CORE and
conventional recovery.  MTTF is determined by three variables: storage
capacity of each node $S$, transfer rate $B$ and node failure rate $\lambda$.
First, we fix the mean failure time $1/\lambda=$ 4 years
\cite{sathiamoorthy13} and $S=$1TB, and evaluate the impact of $B$ on the
MTTFs.  Figure~\ref{fig:mttfanalyze}(a) shows the MTTF results.  With the
increasing transfer rate, the recovery rate and hence the MTTFs of both CORE
and conventional recovery increase.  Next, we fix $B=1$Gbps and $S=$1TB, and
evaluate the impact of $\lambda$ on the MTTFs.
Figure~\ref{fig:mttfanalyze}(b) shows the results.  Both CORE and conventional
recovery see a decreasing MTTF as $\lambda$ increases. 
From both Figures~\ref{fig:mttfanalyze}(a) and \ref{fig:mttfanalyze}(b), 
CORE has a larger MTTF than conventional recovery (by 10-100 times), since
it has a higher recovery rate with less recovery bandwidth.  For example,
considering $T=$1TB, $B=$1Gbps and $\lambda=$0.25, the MTTF of CORE is
26$\times$ of that of conventional recovery.

\section{Implementation}
\label{sec:implementation}

We complement our theoretical analysis with prototype implementation.
As a proof of concept, we implement CORE as an extension to the Hadoop
Distributed File System (HDFS) \cite{shvachko10}.  We modify the source code
of HDFS and its erasure code module HDFS-RAID \cite{hdfsraid}.  We point out
that CORE is also applicable for general large-scale distributed storage
systems. 

\subsection{Overview of HDFS-RAID}
\label{subsec:hdfsraid}

%HDFS-RAID  system designed for reliably storing large datasets in cloud
%computing. 
%We first overview the design architecture of HDFS-RAID, as shown in
%Figure~\ref{figHDFSRAID}.  
By default, HDFS uses 3-way replication to achieve data availability.  To
provide data availability with smaller storage overhead, HDFS-RAID is designed
to convert replicas into erasure-coded data and stripe the erasure-coded data
across different nodes.  We call it the {\em striping} operation. 

HDFS-RAID uses a distributed RAID file system (DRFS) that manages the
erasure-coded data stored in HDFS.  In the original HDFS design, the basic
data unit of the read/write operation is called a block (see
Section~\ref{subsec:basics}).  There are a single NameNode and multiple
DataNodes. The NameNode stores the metadata for HDFS blocks, while the
DataNodes store HDFS blocks.  On top of HDFS,  HDFS-RAID adds a new node
called the RaidNode, which performs the striping operation.  It also
periodically checks any lost blocks, and if needed, performs the recovery
operation for those blocks.  Also, HDFS-RAID provides a client-side interface
called DRFS client, which handles all read/write requests for the
erasure-coded data stored in HDFS.  If a lost block is requested, then it
performs degraded reads to the lost block.  Both the RaidNode and the DRFS
client have an ErasureCode module, which performs the encoding/decoding
operations for the erasure-coded data.  

The striping operation is carried out as follows.  For a given $(n,k)$, the
RaidNode first downloads a group of $k$ blocks (from one of the replicas for
each block).  It then encodes the $k$ blocks into $n$ blocks on a per-stripe
basis (see Section~\ref{subsec:basics}).  The $n$ blocks are then placed on
$n$ DataNodes.  Unused replicas of the $k$ blocks will later be removed from
HDFS.  The RaidNode repeats the same process for another group of $k$ blocks. 

%\begin{figure}[t]
%\centering
%\includegraphics[width=0.9\linewidth]{figs/hdfsraid.eps}
%\vspace{-1em}
%\caption{The architecture of HDFS-RAID.}
%\label{figHDFSRAID}
%\end{figure}

%There are several components in this file system.  The most important part of
%HDFS RAID is {\em RaidNode}, it is a daemon periodically scans the recently
%unmodified files and transfer them from 3-duplicated format to erasure coded
%format thus reduce the storage overhead. RaidNode also periodically checks
%the lost chunks, recomputes and inserts them back to the DRFS.  When the
%administrator wants to manully start the recovery utility, {\em RaidShell}
%can provided this function.  On the client side {\em DRFS client} is a layer
%on top of the HDFS client. When an application send a read request, DRFS
%client directly passes the call to the HDFS client, if HDFS client returns an
%error message, meaning the requested chunk is lost, the DRFS client will
%perform degraded read, reconstruct the lost chunk and return to the
%application.  The encoding/decoding function is provided by {\em
%ErasureCode}, an underlying component. ErasureCode does not only performs
%the calculation, it also retrieves data from data nodes.  On the data
%node, there are a set of signal handlers. These handlers handle the data
%send and receive requests.
%\begin{itemize}
% \item DRFS(distributed raid file system)
% \item RAIDNode(Encode/migration)
% \item BlockFixer(recovery)
% \item ErasureCode(Encode/Decode Interface)
% \item DataNode(a set of Signal Handlers)
% \item DRFSClient(Handling Read Request, both normal and degraded).
%\end{itemize}

\subsection{Integration into HDFS-RAID}
\label{subsec:integration}

To integrate our relayer model into HDFS-RAID, we can simply deploy a relayer
daemon in the RaidNode and the DRFS client for failure recovery and degraded
reads, respectively.  CORE is implemented on HDFS release 0.22.0 with
HDFS-RAID enabled.  We modify both the RaidNode and the DRFS client
accordingly to support concurrent recovery.  Since regenerating codes need
DataNodes to generate encoded symbols during recovery, we add a signal handler
in each DataNode to respond to the request of encoded symbols.  During
recovery, the RaidNode or the DRFS client notifies the surviving DataNodes about
the identities of the failed nodes, and the DataNodes accordingly generate the
encoded symbols. 

{\bf Optimizations of coding.}  In our current prototype, we implement RS
codes \cite{reed60} and IA codes \cite{suh11} as candidates of erasure codes
and regenerating codes, respectively.  We implement them in the ErasureCode
module of HDFS-RAID.  To minimize the computational overhead of
the encoding/decoding operations,  we implement the coding schemes in C++
using the Jerasure library \cite{plank09}, and have the ErasureCode module
execute a specific coding scheme through the Java Native Interface (note that
HDFS-RAID is written in Java).  
For each code we implement, we add {\em XOR transformation} \cite{bloemer95},
which changes all encoding/decoding operations into purely XOR operations, and
{\em XOR scheduling} \cite{hafner05}, which reduces the number of redundant
XOR operations during encoding/decoding.  Both XOR transformation
and XOR scheduling are available in the Jerasure library \cite{plank09}.  
%Furthermore, we implement the XOR operations based on the Streaming SIMD
%Extensions (SSE), an instruction set extension for the x86 architectures.  
%Our evaluation shows that such optimizations greatly improve the
%encoding/decoding performance (see Section~\ref{sec:simulation}). 

\begin{figure}[t]
 \centering
 \includegraphics[width=5in]{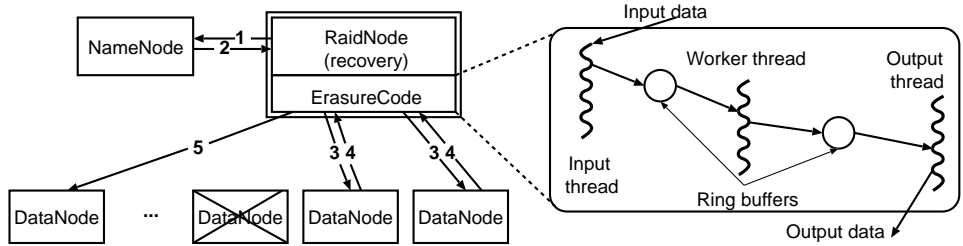}
 \caption{Illustration of the pipelined implementation in CORE for the
	 recovery operation, assuming that we recover a single failure. The same
	 implementation applies to striping (in the RaidNode) and degraded
	 reads (in the DRFS client).} 
 \label{figRecAndPipeline}
\end{figure}

{\bf Pipelined model.}
The original HDFS-RAID uses a single-threaded implementation.  For further
speedup, we implement a {\em pipelined} model that leverages multi-threading
to parallelize the encoding/decoding operations. 
%Our intuition is that modern architectures are built on multi-core
%technologies, so multi-threading can feasibly bring performance gain.
Figure~\ref{figRecAndPipeline} shows the implementation of our pipelined
design in CORE, assuming that a single failure is to be recovered. 
The RaidNode requests metadata from the NameNode (Steps~1-2) and downloads
blocks from the surviving nodes (Steps~3-4).  Then the RaidNode reconstructs
the lost data using the pipelined implementation, which is composed of three
stages.  First, we have an {\em input thread} that collects data from the
surviving DataNodes.  The input thread then dispatches the data via a shared
ring buffer to the {\em worker thread}, which reconstructs the lost data for
the failed nodes.  In the case of regenerating codes, the worker thread
fetches the encoded symbols of one stripe from the ring buffer.  It decodes
the encoded symbols corresponding to the stripe and reconstructs the lost
strips for the failed nodes.  It sends the reconstructed strips to an 
{\em output thread}, and processes another stripe.  The output thread then
collects all reconstructed stripes and uploads the resulting blocks (Step~5).
%We point out that our pipelined model is also applicable for striping
%(performed by the RaidNode) and degraded read (performed by the DRFS client)
%operations. 

\section{Prototype Experiments}
\label{sec:experiment}

We experiment CORE on a distributed storage system testbed.  A major
deployment issue is that the overall recovery performance is determined by a
combination of factors including network bandwidth, disk I/Os,
encoding/decoding overhead.   We address the following questions: 
(i) Does minimizing recovery bandwidth play a key role in improving the
overall recovery performance (see Section~\ref{subsec:microbenchmark})?
(ii) Can CORE preserve the performance of the normal striping operation
offered by HDFS-RAID (see Section~\ref{subsec:striping})?
(iii) How much can CORE improve the performance of recovery, degraded reads,
and MapReduce (see Sections~\ref{subsec:eval_recovery}-\ref{subsec:mapreduce})?

We conduct our experiments on an HDFS testbed with one
NameNode and up to 20 DataNodes being used.  Each node runs on a quad-core
PC equipped with an Intel Core i5-2400 3.10GHz CPU, 8GB RAM, and a Seagate
ST31000524AS 7200RPM 1TB SATA harddisk.  All machines are equipped with a
1Gb/s Ethernet card and interconnected over a 1Gb/s Ethernet switch.  They all
run Linux Ubuntu 12.04. 
%On top of the HDFS testbed, we deploy a relayer on a more advanced quad-core
%PC equipped with an Intel Core i5-760 2.8GHz CPU and 8GB RAM.  The relayer
%acts as either the RaidNode for the striping and recovery operations, or the
%DRFS client for the degraded read operation.  

We compare RS codes \cite{reed60}, which use conventional recovery, and 
CORE, which builds on IA codes \cite{suh11} (see
Section~\ref{subsec:integration}).  Both codes are implemented in C++ and 
compiled with GCC 4.6.3 with the -O3 option.  Our microbenchmark results (see
Section~\ref{subsec:microbenchmark}) are averaged over 10 runs, while other
macrobenchmark results are averaged over five runs. 

%Result
\begin{figure*}[t]
\centering
\begin{tabular}{c@{\ }c@{\ }c}
\includegraphics[width=2.1in]{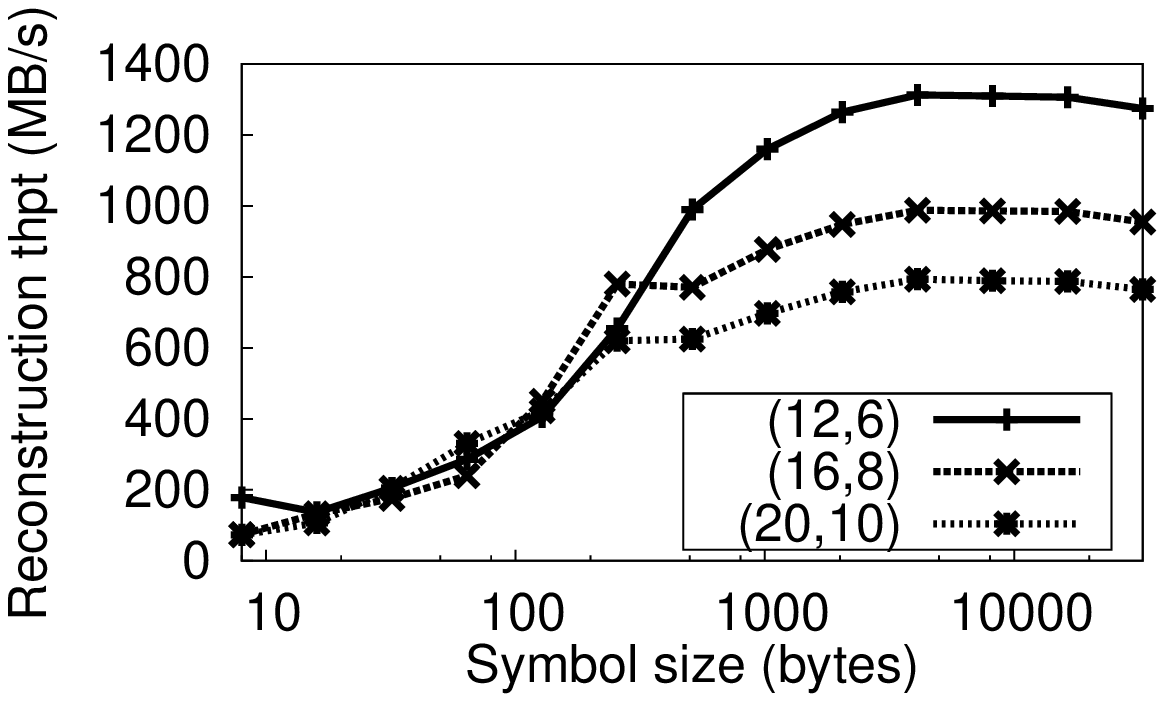} &
\includegraphics[width=2.1in]{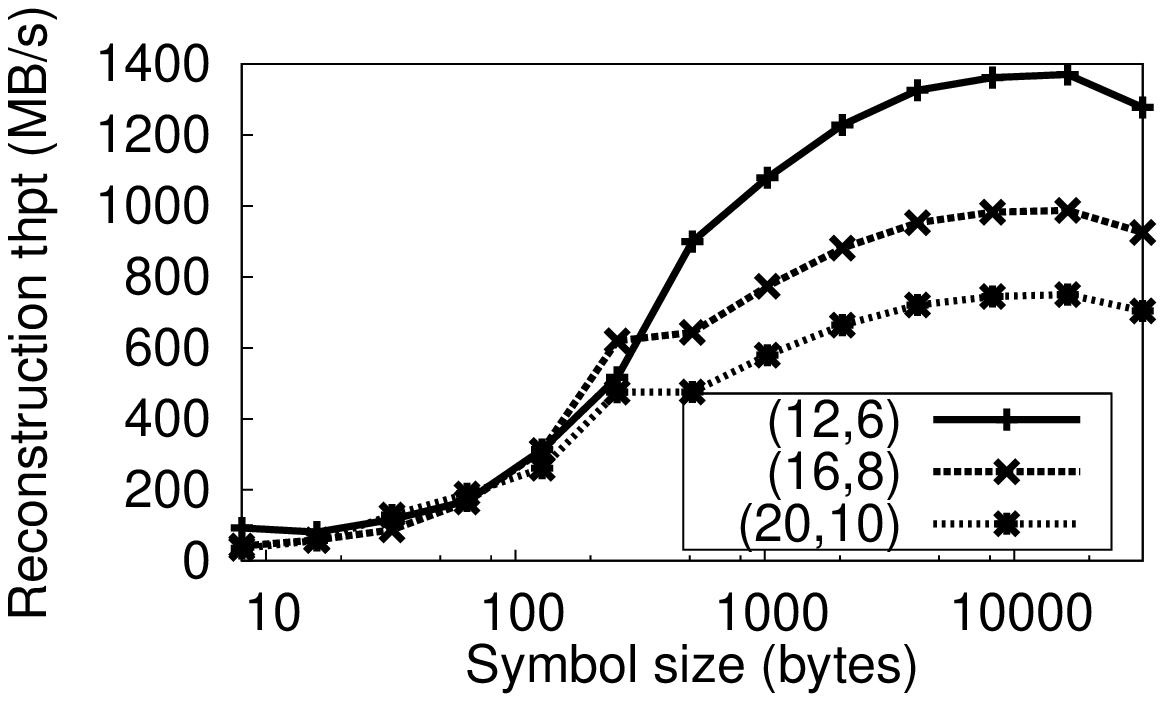} &
\includegraphics[width=2.1in]{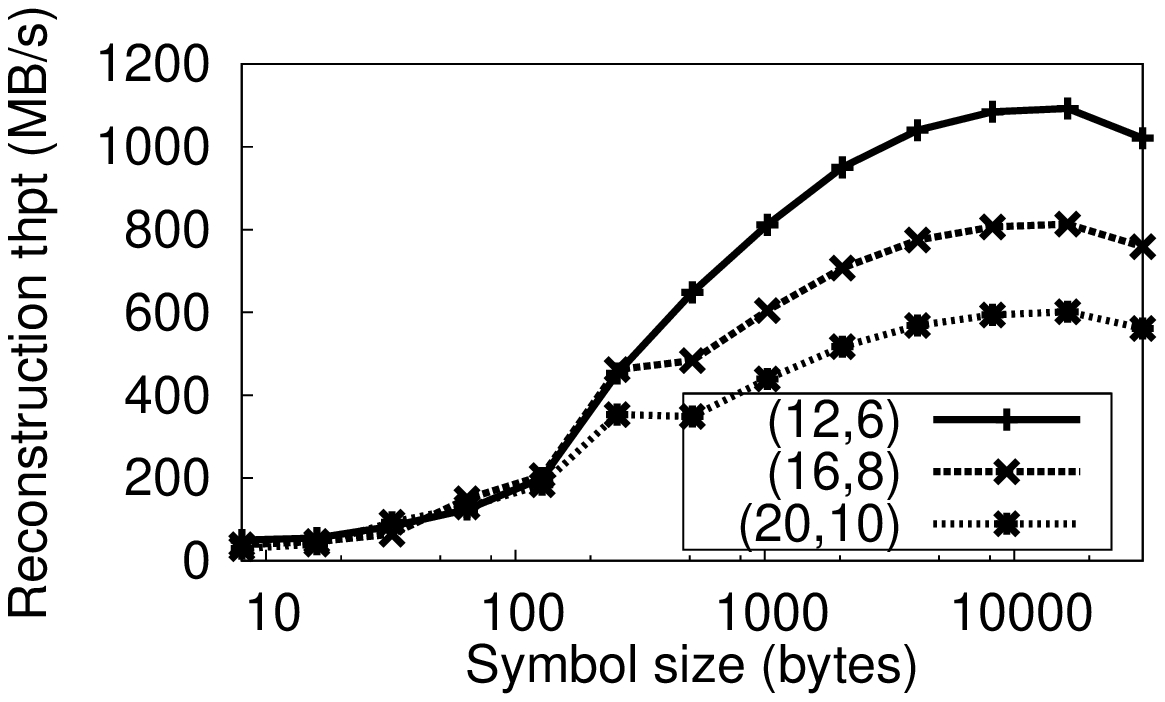} \\
\mbox{(a) RS, $t=1$} & 
\mbox{(b) RS, $t=2$} & 
\mbox{(c) RS, $t=3$} \\
\includegraphics[width=2.1in]{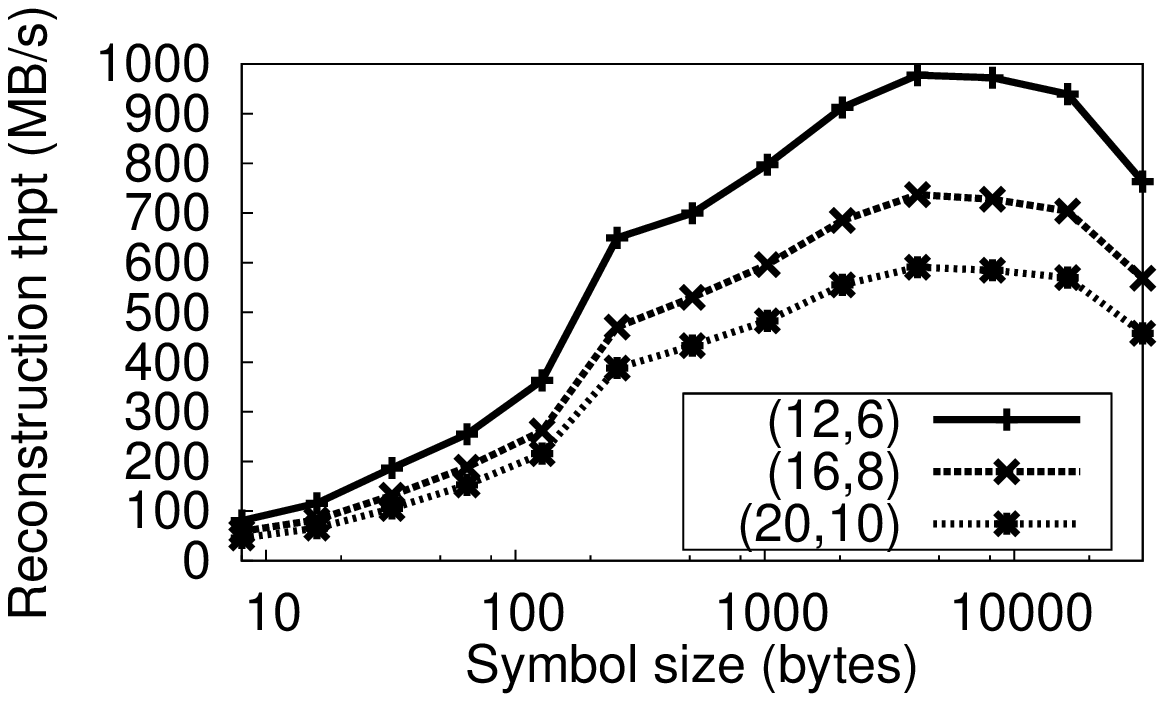} &
\includegraphics[width=2.1in]{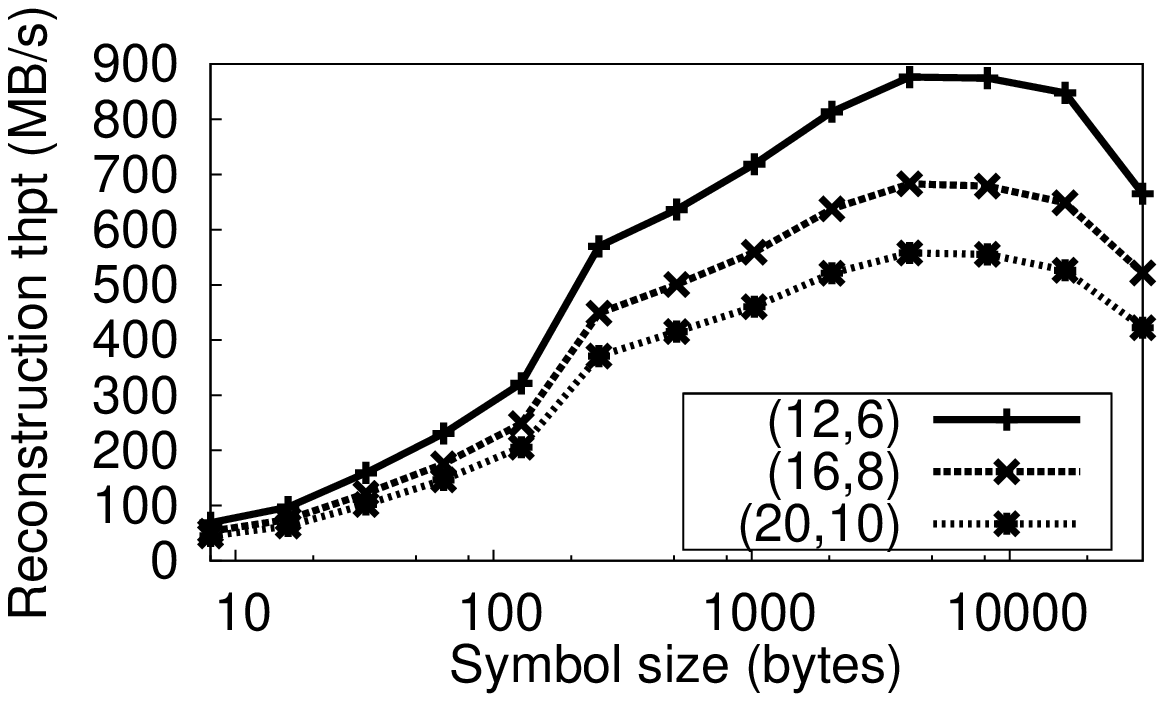} &
\includegraphics[width=2.1in]{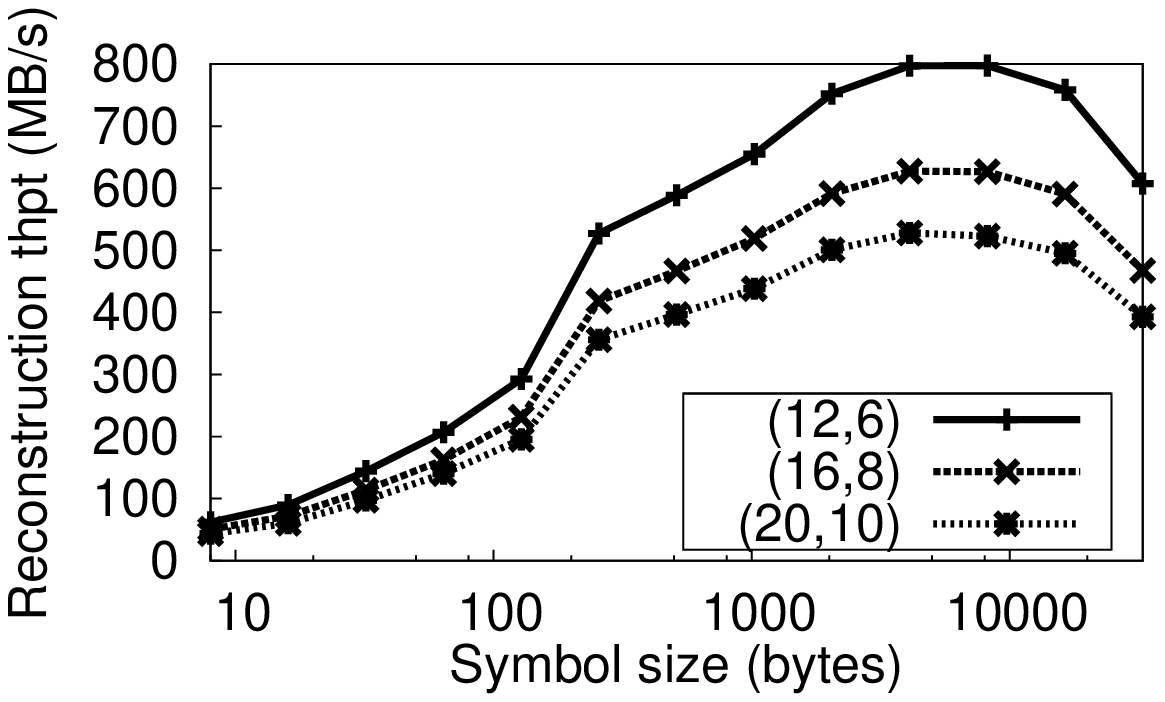} \\
\mbox{(d) CORE, $t=1$} & 
\mbox{(e) CORE, $t=2$} & 
\mbox{(f) CORE, $t=3$} 
%\\
%\includegraphics[width=2.1in]{PMsymbolsize1decode.eps} &
%\includegraphics[width=2.1in]{PMsymbolsize2decode.eps} &
%\includegraphics[width=2.1in]{PMsymbolsize3decode.eps} \\
%\mbox{(g) PM, $t=1$} & 
%\mbox{(h) PM, $t=2$} & 
%\mbox{(i) PM, $t=3$}
\end{tabular}
\caption{Reconstruction throughput of RS codes and CORE versus the symbol size
for different $(n,k)$.} 
\label{fig:symbolsize}
\end{figure*}

\subsection{Microbenchmark Studies}
\label{subsec:microbenchmark}

In this subsection, we conduct microbenchmark studies on the recovery
operation. We first evaluate the encoding/decoding performance versus the
symbol size. We then provide a breakdown analysis on different recovery steps. 

{\bf Encoding/decoding performance in reconstruction.} To evaluate the
computational encoding/decoding overhead of RS codes and CORE in recovery, we
measure how fast the relayer decodes the symbols downloaded from surviving
nodes and reconstructs the lost data.  Since the encoding/decoding operations
are performed over symbols (see Section~\ref{subsec:basics}), our goal here is
to study how the symbol size affects the encoding/decoding performance in
reconstruction. 
%The first one is {\em striping throughput}, which evaluates how fast the
%relayer encodes the original data into encoded data.  It is defined as the
%size of the original data divided by the total encoding time.  The second
%one is 

We vary the symbol size from 8~bytes to 32KB.  Our evaluation operates on 30
stripes of data for different sets of $(n,k)$.  To stress test the
computational encoding/decoding performance, we eliminate the impact of disk
I/Os by first loading the data that is to be downloaded by the relayer for
recovery into memory.  We then measure the time for performing all
encoding/decoding operations on the in-memory data for reconstruction.  We
compute the {\em reconstruction throughput}, which is defined as the size of
the lost data divided by the reconstruction time.

Figure~\ref{fig:symbolsize} shows the reconstruction throughput for one to
three failures for RS codes and CORE.  Larger $(n,k)$ implies more failures
can be tolerated, but has smaller reconstruction throughput since the
generator matrix becomes larger and there is higher encoding/decoding overhead. 
Note that the throughput trend versus the symbol size also conforms to the
results of different erasure codes in the study \cite{plank09}.  The
throughput initially increases with the symbol size, and reaches maximum when
the symbol size is around 4KB to 8KB.  When the symbol size further increases,
the throughput drops because of cache misses \cite{plank09}.  

RS codes have higher reconstruction throughput than CORE (which builds
on IA codes).  The reason is that the strip size of regenerating codes is $r =
n-k$ (see Section~\ref{subsec:regenerating}), while we can implement erasure
codes with $r = 1$.  For the same $(n,k)$, the generator matrix of
regenerating codes is larger than that of erasure codes (see
Section~\ref{subsec:basics}).  Nevertheless, in all cases we consider, CORE
has at least 500MB/s of reconstruction throughput at symbol size 8KB.  Our
following benchmark results show that the reconstruction performance is
{\em not} the bottleneck in the recovery operation. 

%In the following experiments, we fix the symbol size at 8KB, for
%which PM codes achieve a minimum throughput of 64MB/s and IA codes
%achive a minimal throughput of 272 MB/s.

%\begin{figure}
%    \centering
%    \includegraphics[width=0.8\linewidth]{figs/coreRS.eps}
%    \caption{Striping/Reconstruction Throughput of CORE and RS code for $(20,10)$
%    coding schemes}
%    \label{fig:corers}
%\end{figure}

%Note that our
%Jerasure-based implementation of erasure codes can reach a throughput of
%several hundreds of MB/s based on our own evaluation (not shown in the
%figure).  

%Figure \ref{fig:corers} compares the reconstruction bandwidth of CORE and RS
%code for $(20,10)$ coding schemes. For all the parameters, RS code
%outperforms CORE.  For single node failure, the throughput of RS code is as
%high as 1625MB/s, while IA code achieves 571MB/s and PM code achieve 161MB/s.

{\bf Breakdown analysis.}  Recall from Figure~\ref{fig:relayer_model} that a
recovery operation can be decomposed into five different steps.  We now
conduct a simplified analysis on the expected performance of each recovery
step in RS codes and CORE.  Our goal is to identify the bottleneck, and hence
justify the need of minimizing recovery bandwidth. 

We fix the storage capacity of each node to be 1GB.  Suppose that we recover
$t$ failed nodes with a total of $t$GB of data, and that $(n,k) = (20,10)$ is
used.  We collect the system parameters based on the measurements on our
testbed hardware, and derive the expected time for each recovery step as shown
in Table~\ref{tab:corerstime}.  We elaborate our derivations as follows. 
\begin{itemize}
\item
{\em I/O step.}  In both RS codes and CORE, each surviving node reads all its
stored data.  For our disk model, our measurements (using the Linux command
{\tt hdparm}) indicate that the disk read speed is 116MB/s.  Suppose that all
surviving nodes read data in parallel.  In the I/O step, both schemes take
1GB$\div$116MB/s $\approx$ 8.83s. 
\item
{\em Encode step.}  In RS codes, surviving nodes do not perform encoding,
while in CORE, surviving nodes encode their stored data.  Suppose that all
surviving nodes perform the encode step in parallel. Our measurements indicate
that the encoding time on an i5-2400 machine is no more than 0.4 seconds for
1GB of raw data. 
\item
{\em Download step.}  The relayer downloads data from other surviving nodes
via its 1Gb/s interface, so its effective transfer rate must be upper bounded
by 1Gb/s (or 125MB/s).  For RS codes, the relayer always downloads the same
amount of original data, which is $k\times$1GB = 10GB. For CORE, we consider
only the good failure patterns, which account for the majority of cases (see
Section~\ref{subsec:analysis}).  From Theorem~\ref{thm:thm1}, the relayer
downloads $0.1t(20-t)$GB of data (where $t<k=10$).  We can derive the (minimum)
download times for RS codes and CORE accordingly.  In reality, the effective
transfer rate is lower than 1Gb/s and the download times will be higher.
\item
{\em Reconstruction step.}  We fix the symbol size at 8KB, in which both RS
codes and CORE can achieve high reconstruction throughput according to our
previous experiments.  The reconstruction throughput values of RS codes are 
594-789MB/s, while those of CORE are 523-585MB/s.  We derive the
reconstruction times by dividing $t$GB by the reconstruction throughput for
$t$ failures. 
\item
{\em Upload step.}  The relayer uploads $t$GB of reconstructed data via its
1Gb/s interface.  We derive the upload times as in the download step. 
\end{itemize}

From our derivations, we see that the download step uses the most time among
all operations.  Since we can pipeline the download, reconstruction, and
upload steps in the relayer, we can see that the download step is the
bottleneck.  This justifies the need of minimizing recovery bandwidth, which
we define as the amount of data transferred in the download step. 

%\begin{singlespace}
\begin{table}
\centering
\caption{Time comparisons for different recovery steps in RS codes and CORE in
(20,10), assuming 1GB data per node.}
\label{tab:corerstime}
\begin{tabular}{|c||c|c|c|c|c|c|}
	\hline
	\multirow{2}{*}{time(s)} &  RS & RS  &  RS & CORE & CORE   &  CORE \\
				&  $t=1$ & $t=2$   &  $t=3$ &  $t=1$ & $t=2$   &  $t=3$ \\
	\hline
	I/O			&   8.83&  8.83&  8.83&  8.83&  8.83&  8.83 \\
	Encode      &   0&  0&  0&   0.12  &   0.23  &   0.35 \\
	Download    &   81.92 &  81.92&  81.92&   15.56&   29.49&   41.78 \\
	Reconstruct &   1.30&   2.75&   5.17&   1.75&   3.69&  5.87 \\
	Upload      &   8.19&   16.38&   24.58&  8.19&   16.38&   24.58   \\
	\hline
\end{tabular}
\end{table}
%\end{singlespace}

\subsection{Striping}
\label{subsec:striping}

We now evaluate the striping operation that is originally provided by
HDFS-RAID when encoding replicas with RS codes and IA codes (used by CORE).
We also compare our pipelined implementation with the original single-threaded
implementation in HDFS-RAID.  Our goal is to show that CORE, when using IA
codes, maintains the striping performance when compared to RS codes. 

For a given $(n,k)$, we configure our HDFS testbed with $n$ DataNodes, one of
which also deploys the RaidNode.  We prepare a $k$GB of original data as
our input.  By our observation, the input size is large enough to give a
steady throughput.  HDFS first stores the file with the default 3-replication
scheme.  Then the RaidNode stripes the replica data into encoded data using
either RS codes or IA codes.  The encoded data is stored in $n$ DataNodes.  We
rotate node identities when we place the blocks so that the parity blocks are
evenly distributed across different DataNodes to achieve load balancing.  We
fix the symbol size at 8KB.  We use the default HDFS block size at 64MB, but
for some $(n,k)$, we alter the block size slightly to make it a multiple of
the strip size (which is $(n-k)\times$8KB) for IA codes.  We measure the 
{\em striping throughput} as the original size of data divided by the total
time for the entire striping operation.  

%For a given $(n,k)$, we stripe the encoded data into $n$ DataNodes, $k$ of
%which store the original data.  
%We set the Galois Field GF($2^w$) with $w =
%8$, the symbol size at 8KB (see Section~\ref{subsec:encoding}), and the
%default HDFS block size at 64MB.  For some $(n,k)$, we alter the block size
%slightly by 1MB to make it a multiple of the strip size (which is
%$(n-k)\times$8KB) for regenerating codes.  We measure the 
%{\em striping throughput} as the original size of data (i.e., 5GB) divided by
%the total time for the entire striping process.  Our results are averaged over
%five runs. 

\begin{figure}[t]
\centering
\includegraphics[width=4in]{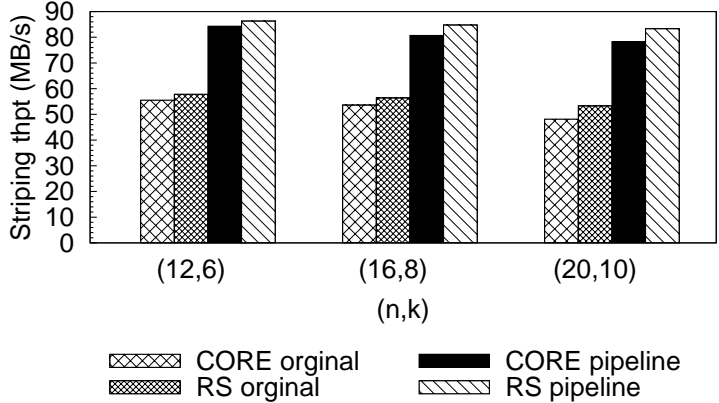}
\caption{Striping throughput.}
\label{fig:striping}
\end{figure}

Figure~\ref{fig:striping} shows the striping throughput results.  By
parallelizing the data transfer and encoding/decoding steps, our pipelined
implementation improves the striping throughput by around 50\% over the
original single-threaded implementation in HDFS-RAID.  We see that IA codes
have smaller striping throughput than RS codes in both implementations.  In
single-threaded implementation, IA codes have higher encoding/decoding
overhead and hence show worse performance.  In pipelined implementation, IA
codes have strip size $r=n-k$ and contain more symbols per stripe than RS
codes with strip size $r=1$.  Our pipelined implementation will not start the
encoding thread until the RaidNode downloads the first stripe of symbols for
each group of $k$ blocks (see Section~\ref{subsec:hdfsraid}).  Thus, RS codes
benefit more from parallelization.  However, the throughput drop in IA codes is
small, by at most 6.1\% only in our pipelined implementation.  

\subsection{Recovery}
\label{subsec:eval_recovery}

We evaluate the recovery performance.  We first stripe encoded data across
DataNodes as in Section~\ref{subsec:striping}.  Then we manually delete
all blocks stored on $t$ DataNodes to mimic $t$ failures, where $t=1, 2, 3$.
Since we rotate node identities when we stripe data, the lost blocks of the
$t$ failed DataNodes include both data and parity blocks.  The RaidNode
recovers the failures and uploads reconstructed blocks to new DataNodes (same
as the failed DataNodes in our evaluation).   Here, we deploy the RaidNode in
one of the new DataNodes.  We measure the {\em recovery throughput} as the
total size of lost blocks divided by the total recovery time.  

\begin{figure}[t]
\centering
\includegraphics[width=4in]{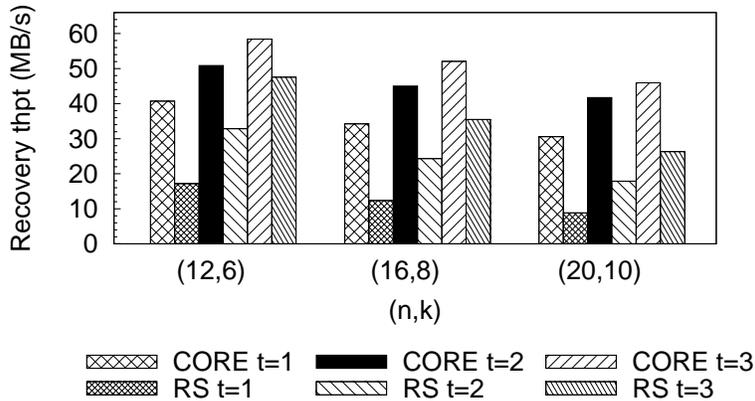}
\caption{Recovery throughput.}
\label{fig:recovery}
\end{figure}

Figure~\ref{fig:recovery} shows the recovery throughput results.  Both RS
codes and CORE see higher throughput for larger $t$ as more lost blocks are
recovered.  Overall, CORE shows significantly higher throughput than RS codes.  
The throughput gain is the highest in (20,10).  For example, for single
failures, the gain is 3.45$\times$; for concurrent failures, the gains are
2.33$\times$ and 1.75$\times$ for $t=2$ and $t=3$, respectively. 

Our experimental results are fairly consistent with our analytical results in
Section~\ref{subsec:analysis}.  For example, in (20,10), the ratio of the
reconstruction bandwidth of CORE to that of erasure codes for $t=2$ and $t=3$
are 0.36 and 0.51, respectively (see Figure~\ref{fig:eval_bandwidth}(a)).
These results translate to the recovery throughput gains of CORE at
2.78$\times$ and 1.96$\times$, respectively.  Our experimental results show
slightly less gains, mainly due to disk I/O and encoding/decoding overheads
that are not captured in the recovery bandwidth. 

\subsection{Degraded Reads}
\label{subsec:eval_degraded}

We further evaluate the degraded read performance in the presence of transient
failures.  The evaluation setting is the same as that of the recovery
operation described in Section~\ref{subsec:eval_recovery}, except that the
degraded read operation is now performed by the DRFS client.  Suppose that $t$
nodes fail, where $t=1,2,3$.  We have the DRFS client request a lost HDFS
block on one of the failed DataNodes.  The lost block will be reconstructed
from the data of other surviving DataNodes.  Here, we deploy the DRFS client
in one of the failed DataNodes.  We measure the {\em degraded read
throughput}, defined as the amount of data being requested divided by the
response time. 

\begin{figure}[t]
\centering
\includegraphics[width=4in]{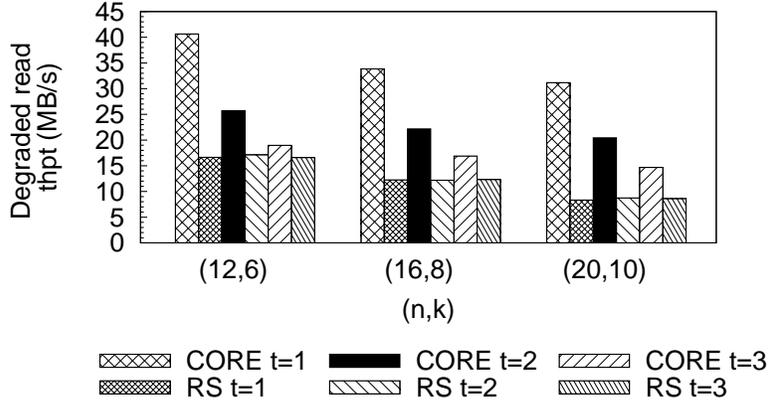}
\caption{Degraded read throughput.}
\label{fig:degradedread}
\end{figure}

Figure~\ref{fig:degradedread} shows the degraded read throughput results.  
RS codes keep almost the same throughput for each $(n,k)$, as they always
download $k$ blocks for reconstruction.  Overall, CORE shows a throughput
gain in degraded reads.  For example, if we consider the (20,10) code, 
CORE shows degraded throughput gain of 3.75$\times$, 2.34$\times$ and 
1.70$\times$ for $t=1$, $t=2$, and $t=3$, respectively. 

We point out that our concurrent reconstruction is optimized for
reconstructing $t$ lost blocks on $t$ failures.  If only one lost block is
reconstructed while $t > 1$, it is possible to use even less reconstruction
bandwidth.  Nevertheless, our results still show the improvements of our
concurrent reconstruction over the conventional one. 

\subsection{Runtime of MapReduce with Node Failures}
\label{subsec:mapreduce}

MapReduce \cite{dean04} is an important data-processing framework running on
top of HDFS.  Here, we conduct preliminary evaluation on how CORE affects the
performance of a MapReduce job with node failures.  

We run a classical WordCount job using MapReduce to count the words in a
document collection.  The WordCount job runs a number of tasks of two types: a
{\em map} task reads a block from HDFS and emits each word to a {\em reduce}
task, which then aggregates the results of multiple map tasks.  With node
failures, some map tasks may perform degraded reads to the unavailable
blocks. 
%There are two types of tasks in the program, {\em map task} and {\em reduce
%task}, and the program contains two phases: {\em map} and {\em reduce}.
%In the map phase, every block of the document is assigned to a map task,
%the map task counts the number of occurrences of each word inside the
%block.  In the reduce phase, for each word, there is a reduce task
%calculating the number of occurrences in the whole document, the reduce
%task receives the number of occurrences in each block from different map
%tasks, sums up the numbers and writes the final result back to HDFS.
%

We consider the same evaluation settings as in
Section~\ref{subsec:eval_degraded}. Here, we focus on (20,10).  We run a
WordCount job on 10GB of plain text data obtained from the Gutenburg website
\cite{gutenberg}.  Using CORE or RS codes, we stripe the encoded blocks across
DataNodes, disable $t$ nodes to simulate a $t$-node failure, and then run the
WordCount job on the encoded data.  We also consider the baseline MapReduce
job when there is no failure.  We use the default MapReduce scheduler to
schedule tasks across DataNodes.  We measure the runtime performance of
different MapReduce components: (i) the average runtime of a normal map task 
running on the available data of normal nodes, (ii) the average runtime a
degraded map task running on the unavailable data of failed nodes, (iii) the
average runtime of a reduce task, and (iv) the overall runtime of the
WordCount job. 

%We measure {\em run-time}, defined as the time required to 
%finish the MapReduce job.  
%We compare the run-time of CORE and RS codes ($t=1,2,3$) with the
%baseline run-time, which is the run-time without node failures. 

%\begin{figure}[t]
%\centering
%\includegraphics[width=0.5\linewidth]{figs/mapreduce.eps}
%\caption{Run-time of MapReduce jobs.}
%\label{fig:mapreduce}
%\end{figure}

\begin{table}
    \centering
	\caption{Runtime (in seconds) of different MapReduce components using RS
		codes and CORE.}
    \label{tbl:mapreduce}
    \begin{tabular}{|c|c||c|c||c|c||c|c|}
        \hline
        & \multirow{2}{*}{baseline}&   CORE & RS & CORE & RS & CORE & RS \\
        &&   $t=1$ & $t=1$ & $t=2$ & $t=2$ & $t=3$ & $t=3$ \\
        \hline
        map task (normal) & 20.48 & 20.43 & 20.43 & 20.46& 20.44& 20.55& 20.54 \\
        map task (degraded) & NA & 23.39& 32.83& 24.23& 31.30& 27.19& 33.14\\
        reduce task & 30.10 & 31.21& 31.39& 31.42& 31.12& 31.25& 30.73 \\
        overall & 209.20 & 212.40& 216.20& 216.60& 231.20& 231.00& 242.80\\
        \hline
    \end{tabular}
\end{table}

Table~\ref{tbl:mapreduce} shows the runtime of different MapReduce components.
For the normal map tasks and the reduce tasks, their runtimes are almost
identical to the baseline, meaning that CORE does not have adverse effects to
such tasks.  The degraded map task incurs a longer time than the baseline due
to degraded reads.  Nevertheless, CORE outperforms RS codes in this item.  For
$t=$1, 2 and 3, CORE takes 29\%, 22\% and 18\% less time than RS codes to run
a degraded map task.  The results also conform to our theoretical findings. 
The extra runtime of the degraded map task over the normal map task is mainly
due to the degraded read request.  Consider $t=1$.  For RS codes, the extra
runtime is 12.4s, while for CORE, the extra runtime is 2.96s (or 80\% less).
This is consistent with our analysis results in Section~\ref{subsec:analysis}. 
	
CORE also improves the overall runtime of the WordCount job, although the
improvement is less significant due to other overheads.  On the other hand, we
expect that the improvement of CORE becomes more significant in a large-scale
distributed setting where network bandwidth is limited.  We argue that the
MapReduce evaluation here is preliminary.  We plan to consider more workloads
and testbed environments in future work. 

%Figure~\ref{fig:mapreduce} shows the MapReduce run-time experiment results.
%Compared with the baseline result, both CORE and RS codes have longer
%run-time due to the degraded read operations.  Overall, the run-time of
%CORE is less than the run-time of RS codes.  For example, consider (20,10), 
%for $t=1,2,3$, the run-time results of CORE are 6.2\%, 30.5\% and  
%62.0\% longer than the baseline results, respectively, while for RS codes, 
%the run-time results are 14.3\%, 47.5\% and 72.4\% longer.  
%
%We note that for MapReduce run-time experiment, the improvement of CORE 
%over RS codes is not as great as degraded read experiment.  
%For example, (20,10) and $t=1$, for the degraded read experiment, 
%CORE takes 73.3\% less time for a degraded read request than RS codes.  
%While for MapReduce experiments, the extra run-time for CORE is 
%56.6\% less than RS codes.  The reason for less improvement is that
%the mapper will first try to read the lost block, and trigger the
%degraded read function when the read request is timeout, this
%timeout increases the overhead of MapReduce job.  Also, CORE involves
%more calculation than RS code (i.e., node encoding and more complex
%recovery function), thus affects the run-time of MapReduce job.

\section{Related Work}
\label{sec:related}

We review related work on the recovery problem for erasure codes and
regenerating codes. 

%{\bf Efficient recovery.}  There have been extensive studies on improving the
%recovery performance of coded storage systems.  Muntz and Lui \cite{muntz90}
%use parity declustering (i.e., distributing data over a larger number of
%disks) to reduce performance degradation during recovery.  Holland 
%{\em et al.} \cite{holland94} analyze the impact of parity declustering, and
%further propose workflow parallelization to speed up reconstruction.  FARM
%\cite{xin04} employs declustering to improve both recovery and reliability in
%large-scale storage systems.  Total Recall \cite{bhagwan04} proposes a lazy
%repair scheme to reduce the data transferred when recovering concurrent
%failures.  Some studies \cite{sivathanu05,tian07,wu09} leverage workload
%characteristics or access patterns to improve the reconstruction performance.
%Note that all the above approaches build on our definition of conventional
%recovery, in which first reconstruct the original data.  They retrieve the
%size of original data, and incur high I/Os and bandwidth in general. 

{\bf Minimizing I/Os.}  Several studies focus on minimizing I/Os required for
recovering a single failure in erasure codes.  Their approaches mainly focus
on a disk array system where the disk access is the bottleneck.  Authors of
\cite{wang10,xiang11} propose optimal single failure recovery for RAID-6
codes.  Khan {\em et al.} \cite{khan12} show that finding the optimal recovery
solution for arbitrary erasure codes is NP-hard.
%and propose an enumeration-based recovery algorithm.  They also propose a
%modified Reed-Solomon code for efficient degraded reads.  Authors of
%\cite{zhu12a,zhu12b} propose greedy heuristics to speed up the search of
%solutions for single failure recovery.  
Note that the performance gains of the above solutions over the conventional
recovery are generally less than 30\%, while regenerating codes achieve a much
higher gain in single failure recovery (see Section~\ref{sec:experiment}).  

%Huang {\em et al.} \cite{huang12} propose local recovery codes that reduce the
%bandwidth and I/O when reconstructing a lost data fragment. They evaluate the
%codes atop the Windows Azure Storage system.  Sathiamoorthy {\em et al.}
%\cite{sathiamoorthy13} also propose local recovery codes, and evaluate the
%codes atop HDFS-RAID as in our work.  

Authors of
\cite{huang12,esmali13,papailiopoulos12,sathiamoorthy13}\footnote{Although the
proposed scheme of \cite{esmali13} is also called CORE, it refers to Cross
Object Redundancy and builds on local recovery codes, which have very
different constructions from regenerating codes considered by our work.} have
proposed local recovery codes that reduce bandwidth and I/O when recovering
lost data.  They evaluate the codes atop a cloud storage system simulator 
(e.g., in \cite{papailiopoulos12}), Azure Storage (e.g., in \cite{huang12})
and HDFS (e.g., in \cite{esmali13,sathiamoorthy13}).
It is worth noting that the local recovery codes are non-MDS codes with
additional parities added to storage, so as to trade for better recovery
performance.  All these studies focus on optimizing single failure recovery.
Our work differs from them in several aspects: (i) we consider optimal minimum
storage regenerating codes that are MDS codes, (ii) we consider recovering
both single and concurrent failures, (iii) we experiment regenerating codes
that require storage nodes to perform encoding operations. 

{\bf Minimizing recovery bandwidth.} Regenerating codes \cite{dimakis10} 
minimize the recovery bandwidth for a single failure in a distributed storage
system.  There have been many theoretical studies on constructing regenerating
codes (e.g., \cite{dimakis10,rashmi09,rashmi11b,shah12,suh11}).  In contrast
with the above solutions that minimize I/Os, most regenerating codes typically
read all stored data to generate encoded data.  Implementation studies of
regenerating codes recently receive attention from the research community,
such as \cite{duminuco09,hu12,huang11,jiekak12}.  Note that the studies
\cite{duminuco09,huang11,jiekak12} do not integrate regeneration codes into a
real storage system, while  NCCloud \cite{hu12} implements a storage
prototype based on non-systematic regenerating codes.  
%We point out that existing implementation studies only focus on single
%failure recovery. 

{\bf Cooperative recovery.} Several theoretical studies (e.g.,
\cite{hu10,kermarrec11,shum11a,shum11b}) address concurrent failure
recovery based on regenerating codes, and they focus on recovery of lost
data on new nodes.  They all consider a {\em cooperative model}, in which the
new nodes exchange among themselves their data being read from surviving
nodes during recovery.  Authors of \cite{hu10,kermarrec11} prove that the
cooperative model achieves the same optimal recovery bandwidth as ours,  but
they do not provide explicit constructions of regenerating codes that achieve
the optimal point.  Authors of \cite{shum11a,shum11b} provide such explicit
implementations, but they focus on limited parameters and the resulting
implementations do not provide any bandwidth saving over erasure codes.  A
drawback of the cooperative model requires coordination among the new nodes to
perform recovery, and its implementation complexities are unknown.  Extending
it for degraded reads is also non-trivial, as clients simply request lost data
instead of recovering lost data on new nodes.

\section{Discussion}
\label{sec:discussion}

In this section, we discuss several open issues that are not covered in this
paper.

{\bf High redundancy of CORE.} In this paper, we consider the MSR codes with
fairly high redundancy (i.e., double redundancy), due to the requirements
imposed by the underlying constructions of optimal exact regenerating codes.
It is shown in \cite{shah12} that all $(n,k)$ linear MSR codes with exact
recovery must satisfy the condition $n\ge 2k-2$.  Other $(n,k)$ codes may be
constructed via the non-systematic, functional regenerating codes
\cite{dimakis10}, which are suited to the rarely-read data.  How to extend
CORE for functional regenerating codes remain an open issue in this work. 
%achieve lower bound for node failure recovery with functional recovery.
%Functional recovery means that the regenerated data is not exactly the same
%as the lost data, but the MDS property of the system is kept.  The idea of
%CORE can be extended to support codes of low redundancy with functional
%recovery, but we have not looked into this problem yet.

{\bf Concurrent recovery of non-MDS codes.} We consider the concurrent
recovery problem of MSR codes, which achieve the minimum storage efficiency 
as in MDS codes (see Section~\ref{subsec:basics}).  One may consider the
non-MDS codes, which incurs higher storage overhead but achieve better single
failure recovery performance (e.g. MBR codes \cite{rashmi09} and local
recovery codes \cite{huang12,esmali13,papailiopoulos12,sathiamoorthy13}). An
open issue is how to extend these non-MDS codes to support efficient
concurrent recovery. 

{\bf Wide-area storage systems.}  We currently implement CORE atop HDFS.  We
plan to explore the implementation of CORE in wide-area storage systems (e.g.,
\cite{bhagwan04,cleversafe,dabek01,kubiatowicz00}), where network bandwidth is
limited and the benefits of regenerating codes should become more prominent. 
Also, one side benefit of CORE is that we can delay recovery until the number
of failed nodes reaches some threshold so as to we avoid recovering transient
failures that are commonly found in wide-area networks
\cite{bhagwan04,chun06,nath06}. 

%Also, recovery operation degrades system performance, delaying recovery can
%reduce the impact on the performance on the whole system.  By taking
%advantage of CORE, we can reduce the recovery time, thus reduce the impact on
%the performance of the whole system, CORE also provides good degraded read
%performance, thus reduce the response for time for requesting unavailable
%data.

\section{Conclusions}
\label{sec:conclusion}

We address the reconstruction problem in a distributed storage system in the
presence of single and concurrent failures, from both theoretical and applied
perspectives.  We explore the use of regenerating codes (or network coding) to
provide fault-tolerant storage and minimize the bandwidth of data transfer
during reconstruction.  We propose a system CORE, which generalizes existing
optimal single-failure-based regenerating codes to support the recoveries of
both single and concurrent failures.  We theoretically show that CORE
minimizes the reconstruction bandwidth in most concurrent failure patterns.
Our scheme adopts a relayer model that can be easily integrated into real
storage systems.  To demonstrate, we prototype CORE as a layer atop Hadoop
HDFS, and show via testbed experiments that we can speed up both recovery and
degraded read operations.  The source code of our CORE prototype is available
for download at: {\bf \url{http://ansrlab.cse.cuhk.edu.hk/software/core}}.

\bibliographystyle{abbrv}
\bibliography{reference}

\section*{Appendix}

\subsection*{Proof of Theorem~\ref{thm:thm1}}

We can formally build our proof based on the analysis of
the information flow graph as in \cite{dimakis10}.  Here, we only show the main
idea.  Let $d$ be the number of surviving nodes from which the relayer
downloads data for recovery. Let $\beta$ be the amount of data downloaded (per
stripe) from each of the $d$ surviving nodes to recover $t$ failed nodes.  We
assume that the reconstructed data will be stored on $t$ new nodes, which
contain a total of $d\beta$ units of information.  

We first consider $t < k$.  Due to the MDS property, we can restore the
original data from any $k$ out of $n$ nodes, each storing $\frac{M}{k}$ units
of data. For example, we can select a set of any $k-\hat{t}$ originally
surviving nodes (denoted by set $\SX$) and a set of any $\hat{t}$ new nodes
(denoted by set $\SY$) for some $\hat{t} \le t$.  The total amount of useful
information must be at least $M$ in order for the original data to be
restorable.  However, $\SY$ contains $(k-\hat{t})\beta$ units of information
derived from $\SX$.  By excluding the redundant information, we require:
\[
\frac{M}{k}(k-\hat{t}) + (d\beta - (k-\hat{t})\beta) \ge M, \textrm{ for any
$\hat{t} \le t$}. 
\]
The left side is minimum when $\hat{t} = t$.  Thus, the recovery
bandwidth (i.e., $d\beta$) must be at least 
$\frac{M\times d\times t}{k(d - k +t)}$.  To minimize the recovery
bandwidth with respect to $d$, we set $d = n-t$ and the result follows. 

When $t\ge k$, any $k$ out of the $t$ new nodes must be able to restore the
original data due to the MDS property.  Thus, the $t$ new nodes must contain
$M$ units of useful information, which can be reconstructed by downloading data
from any $k$ surviving nodes as in erasure codes.  The recovery
bandwidth is $M$.  \done

\subsection*{Proof of Theorem~\ref{thm:thm2}}

Since MSR codes achieve the lower bound of
recovery bandwidth for single failure recovery, the amount
of data downloaded from {\em each} surviving node is $\frac{M}{k(n-k)}$
\cite{dimakis10} (see Equation~(\ref{eqn:msr})). 
	
Consider $t < k$.  CORE in essence performs $t$ single failure recoveries based
on MSR codes, and in each recovery we actually download $\frac{M}{k(n-k)}$
units of data from each of the $n-t$ surviving nodes.  If the failure pattern
is good, then we can recover the virtual symbols and hence the lost data.  The
lower bound is hit for $t < k$.  For $t \ge k$, we can simply download $M$
units of data from any $k$ surviving nodes and {\em any} failure pattern can
be recovered.  The result follows.  \done

\end{document}